\documentstyle[aaspp4]{article}

\def\stacksymbols #1#2#3#4{\def\theguybelow{#2}
	\def\verticalposition{\lower#3pt}
	\def\spacingwithinsymbol{\baselineskip0pt\lineskip#4pt}
	\mathrel{\mathpalette\intermediary#1}}
\def\intermediary #1#2{\verticalposition\vbox{\spacingwithinsymbol
	\everycr={}\tabskip0pt
	\halign{$\mathsurround0pt#1\hfil##\hfil$\crcr#2\crcr
		\theguybelow\crcr}}}
\def\lta{\stacksymbols{<}{\sim}{2.5}{.2}}
\def\gta{\stacksymbols{>}{\sim}{3}{.5}}

\begin{document}
\title{A COMPARISON OF METAL ENRICHMENT HISTORIES IN RICH CLUSTERS
AND INDIVIDUAL LUMINOUS ELLIPTICAL GALAXIES}


\author{Fabrizio Brighenti$^{2,3}$ and William G. Mathews$^2$}

\affil{$^2$University of California Observatories/Lick Observatory,
Board of Studies in Astronomy and Astrophysics,
University of California, Santa Cruz, CA 95064\\
mathews@lick.ucsc.edu}

\affil{$^3$Dipartimento di Astronomia,
Universit\`a di Bologna,
via Zamboni 33,
Bologna 40126, Italy\\
brighenti@astbo3.bo.astro.it}






\vskip .2in

\begin{abstract}

The large spatial extent of hot, X-ray emitting 
gaseous halos around massive 
elliptical galaxies indicates that most of this gas has not 
been generated by stellar mass loss.
Instead, much of this gas results from an intergalactic 
gaseous inflow toward the overdensity from which giant 
ellipticals and their associated galaxy groups formed.
Since these hot gaseous halos are old, they contain 
important information about the star formation history 
of elliptical galaxies.
In this paper we show that the enrichment history of this 
hot gas is closely linked to its gas dynamical history; 
supernovae provide both energy and metal enrichment.
We find that Type II supernovae based on a Salpeter IMF, 
plus a small number of additional Type Ia supernovae,
can explain the the density,
temperature and abundance profiles currently observed 
in gaseous halos around massive ellipticals.
Within the central, optically bright region 
of luminous ellipticals, approximately half of 
the interstellar iron 
is produced by Type Ia supernovae and half 
by mass lost from evolving stars which were 
originally enriched in iron by Type II supernovae.
However, iron and silicon 
abundances in the intracluster gas within rich clusters
suggest enrichment by a larger number of 
supernovae per unit optical light than we require 
for massive ellipticals.
The additional supernovae implied by cluster data 
cannot be reconciled with our models for 
individual massive ellipticals. 
Evidently, rich clusters
cannot be constructed by simply combining
ellipticals and their associated groups since 
the enrichment history of clusters and massive ellipticals
appears to be fundamentally different.
Neither currently discussed resolution of this 
discrepancy  -- increased number of Type II supernovae 
(flat IMF) or strong Type Ia enrichment in 
clusters -- is attractive.
Although the global hot gas iron abundance 
is similar in all large galaxy clusters, 
silicon is enhanced in hotter, richer clusters.
This Si/Fe variation implies that E and S0 galaxies are
not the only sources of cluster gas enrichment; 
perhaps spirals or low mass galaxies are also important.

\end{abstract}



\section{INTRODUCTION}

Luminous elliptical galaxies are the astronomical analog of tree rings or
geological core samples -- within the hot gas in these massive
galaxies resides
important information about the earliest stages of galaxy
formation and the enrichment of the cosmos with heavy elements
from the first generations of stars.
The stellar density profile and low rotation of bright ellipticals
indicate that they were assembled from smaller 
galactic objects containing stars.
The most likely sites for the formation of 
giant elliptical galaxies are small groups 
of galaxies where mutual tidal interactions 
are strong and the likelihood of galactic merging 
is enhanced.
In an idealized version of this formation hypothesis,
the merging process within the group comes to completion after 
a few dynamical times when the massive central elliptical has 
consumed the dark and luminous matter of many 
original group galaxies of moderate mass.
When this happens, the merging process shuts down 
and the elliptical evolves passively 
as a group-dominant elliptical 
in the field or enters a large cluster.
While there is clear evidence that some massive ellipticals contain
younger stars and must have suffered significant mergers 
since the epoch of most intense galaxy formation, 
there is also evidence that other more venerable ellipticals 
have survived until the present time,  
relatively undisturbed by events that have occurred since their 
formation at early times.
The hot interstellar gas in these old galaxies is most interesting 
since it still contains information about the earliest stages of 
star formation, including the heating and enrichment of the gas 
by supernovae.

In this paper we investigate the production of
iron and silicon by Type Ia and 
Type II supernovae (SNIa and SNII respectively) and follow the 
dynamical redistribution of these metals 
in the hot interstellar and intragroup gas until the present time.
Gas enrichment and dynamical evolution are linked 
since the collective energy released by 
Type II supernovae can expel metal-enriched 
gas from elliptical galaxies in a wind.
The energetics of Type II supernovae 
during early star formation 
can be constrained by comparing heavy metal abundances expected 
from evolutionary gas dynamical calculations with 
abundances provided by X-ray observations of individual 
massive ellipticals.
In a similar fashion, metal abundances observed in 
rich clusters of galaxies provide global enrichment 
constraints on early type galaxies which are generally 
regarded as the principal source of metals 
in the intracluster gas.
Rich clusters are thought to have retained all of the 
products of supernova enrichment either 
within galactic stars or in the hot intracluster medium.
In this sense rich clusters are closed boxes.
From an observational standpoint 
individual massive elliptical galaxies are 
certainly not closed boxes 
since most of the metals produced by supernovae 
have been carried out by supernova-driven winds. 
Much of these metals 
now resides in low density gas at large distances 
from the optical galaxy where detection is difficult 
or where the metal enriched 
gas has been tidally or dynamically removed. 
From the standpoint of our gas dynamical models, however, 
individual ellipticals {\it are} closed boxes since we can 
accurately follow the products of supernova enrichment 
into distant regions of very low density.

Thermal X-ray emission from hot circumgalactic gas 
surrounding many bright ellipticals often extends far beyond 
the optical image of the galaxy (Mathews \& Brighenti 1998a).
The large mass and extended spatial distribution 
of this gas cannot be understood only from 
normal evolutionary stellar mass loss.
Our dynamical models indicate that the most distant 
gas in these galactic halos has accumulated by 
secondary infall into the perturbation that
initially led to the formation of the group and
its dominant elliptical
(Brighenti \& Mathews 1998a, 1998b;
Mathews \& Brighenti 1998b).

The objective of our recent gas dynamical calculations 
has been to follow the evolution of the hot gas within  
and around massive elliptical galaxies since the time
of galaxy formation.
A successful model can reproduce the radial 
density and temperature distributions as determined 
from X-ray observations.
In our recent models we begin with an 
overdensity perturbation 
in a simple flat cosmology.
As dark matter accumulates in the core of the
growing perturbation,
we assume that a stationary NFW dark matter halo
is formed (Navarro, Frenk and White 1996).
Exterior to this stationary
halo the dark matter inflow is identical to
the cosmic similarity flow of Bertschinger (1985).
The baryonic gas evolves in this time-dependent
potential: flowing inward, shocking at some
radius and radiatively cooling near the center.
At some early time, $t_*$, when enough baryons
have accumulated in the perturbation,
we form the de Vaucouleurs profile of the large
elliptical observed today.
By this means we circumvent the complex merging processes
that occurred at very early times.
Nevertheless, our calculation globally conserves
dark and baryonic mass and treats the 
gas dynamics in full detail from $t_*$ to the present time.
In these simple models we release 
the Type II supernova  
energy at the same time $t_*$ that
the stellar system is formed but the Type Ia energy 
is released from $t_*$ to the present according to 
some assumed variable rate.

However, there is a degeneracy in our successful solutions. 
We find that the hot gas density and 
temperature profiles typical of massive ellipticals, 
$n(r)$ and $T(r)$, can be reproduced 
with different combinations of the fundamental model parameters:
the time of star formation 
$t_*$, the total SNII energy released $E_{II}$,
and the universal baryon mass fraction 
$\Omega_b \approx 0.05 \pm 0.01$
as determined by big bang nucleosynthesis (Walker et al. 1991).
For example, solutions with $\Omega_b$ decreased by $\sim 0.01$
are similar to those with $E_{II}$ increased by $\sim 2$ or
with $t_*$ decreased by $\sim 1$ Gyr.

Much of the degeneracy involving 
these uncertain parameters 
can in principle be removed by considering 
the abundance and 
of iron and other elements produced in supernovae.
The total mass of heavy elements in stars and gas today
is directly related to the total number of supernovae of each 
type that have occurred.
Unfortunately, many of the essential supernova-related 
parameters are poorly known: the average amount of 
iron or silicon created in each supernova event, 
the total mass of these elements locked into stars today, 
the initial mass function (IMF), etc.

While we cannot attempt here a definitive theoretical 
resolution of the enrichment history of early type galaxies,
we can make progress within the limitations of current 
theory and observational data.
In view of the uncertainties involved, however, 
a highly detailed gas dynamical calculation 
including the effects of metal enrichment seems unwarranted.
The approximate results 
we present here can be regarded as the first step
in an iterative procedure that will become 
better defined in the future.

Nevertheless, the metal enrichment 
history of elliptical galaxies that we 
describe here is sufficient to 
rule out several scenarios that have been widely discussed.
One of our motivations for beginning this project
is the apparent dissimilarity of iron abundance
and gas fraction between galaxy groups
and rich clusters which led David (1997) and 
Renzini (1997) to remark
that present-day, rich clusters cannot be assembled from
present-day groups.
For example, most (50\% - 80\%)
of the mass of heavy metals in rich clusters resides 
in the intracluster gas 
where the iron mass per unit of stellar light 
$M_{Fe}/L_B$ consistently 
exceeds that in small groups or individual massive ellipticals.
Moreover, the lower fraction of gas mass in galaxy 
groups relative to clusters 
indicates that much of the metal-enriched gas produced 
in groups or individual ellipticals has been expelled 
into the environment by the release of supernova energy
(Arnoud, et al. 1992; David, Jones, \& Forman 1995).
Perhaps, therefore, the gas and iron formerly 
within large ellipticals and galaxy groups 
has been expelled by supernova-driven winds 
into the low density environment where it 
escapes X-ray detection.
However, 
this low density, metal-enriched gas can be followed in 
our gas-dynamical models.
In successful models the metal rich gas remaining 
within the galaxy must agree with the gas abundance and 
abundance gradients observed in these bright galaxies.

Our results reported here indicate, however, that the 
mass of metals that has flowed out of luminous ellipticals
into their low density environment is insufficient to
account for the larger mass of metals observed in 
rich clusters.
We are therefore faced with 
a fundamental inconsistency: parameters that allow the 
intracluster gas to be enriched by galactic winds from massive 
ellipticals are inconsistent with detailed X-ray observations 
of the metal content within individual ellipticals.

The relative importance of SNII and SNIa in creating 
the metals observed in cluster gas has been widely
debated.
The total iron abundance in rich clusters is too large to be 
produced by SNII with a standard Salpeter IMF.
Many authors have suggested that 
SNIa produce enough iron to account for the additional iron 
required (e.g. Ishimaru \& Arimoto 1997; Renzini 1997; Wyse 1997).
This solution is appealing since approximately $3 \over 4$ of the 
iron in our Galaxy is produced by Type Ia supernovae 
(Renzini 1997).
However, we show here that the total silicon produced by both 
types of supernova, when considered along with 
additional gas dynamical constraints,  
indicates that SNIa produce at most only a few percent of the 
total iron in massive ellipticals.
By contrast, if SNII are the primary source of metal enrichment 
in rich clusters then the IMF must be flatter than Salpeter;
this solution to the iron problem has been suggested by 
David (1997) and Gibson, Loewenstein \& Mushotzky (1997).
However, the iron abundance and its radial gradient  
observed in massive ellipticals cannot be fit with 
our gas-dynamical models if the number of SNII exceeds the 
Salpeter value.
It would appear therefore that massive ellipticals may not 
be the primary source of enrichment of the intracluster gas.

We begin our discussion with a review and reanalysis of the observed 
iron and silicon abundances in rich clusters and 
discuss the implications of
these abundances for clusters and individual early type galaxies.

\section{SUPERNOVA ENRICHMENT IN RICH CLUSTERS}

The mean iron abundance observed in the hot gas in rich clusters
is about 0.40 in solar units with a cosmic scatter
(0.3 - 0.65) that exceeds the errors of measurement (Mushotzky 1998).
These values are somewhat greater than the iron abundances quoted
in the X-ray literature, most of which are based on
the ``photospheric'' solar iron abundance (by mass)
$z^{(ph)}_{~~Fe,\odot} = 2.66 \times 10^{-3}$ since we assume here the
lower, ``meteoritic'', solar iron abundance
$z_{Fe,\odot} = 1.83 \times 10^{-3}$.
This distinction between ``meteoritic'' and ``photospheric''
solar iron abundance was made in the review by
Anders \& Grevesse (1989).
However, recent atmospheric models of the sun indicate that the iron
abundance in the solar photosphere is similar to
the so-called ``meteoritic'' value of Anders \& Grevesse
(Holweger et al. 1991; Biemont et al. 1991; McWilliam 1997).
The implications of
the lower $z_{Fe,\odot}$ for X-ray abundances has been discussed by
Ishimaru \& Arimoto (1997) and
Gibson, Loewenstein, \& Mushotzky (1997).

Arnaud et al. (1992) have shown that the total gas 
mass in the hot intracluster medium (ICM) of rich clusters 
${\cal M}_g$ and the mass of iron in that gas are 
proportional to the total optical luminosity ${\cal L}_B$
of all early type galaxies in the cluster.
Arnaud et al. find that ${\cal M}_g$ is uncorrelated with 
the total ${\cal L}_{B,sp}$ 
from spiral galaxies in these clusters. 
Based on these findings, 
it is generally assumed that early type galaxies 
dominate the cluster gas enrichment.

The relative contributions of
Type II and Type Ia supernovae
(SNII and SNIa) to the ICM metal enrichment 
has been widely discussed. 
The total mass of a particular heavy element in the ICM 
can be determined from 
the mass of that element produced by supernovae 
and the total number of supernova events in the past.
The number of SNII events can be inferred from the 
total spheroidal stellar mass in the cluster,
together with an assumed initial mass function (IMF) 
to determine the fraction of stars that become 
Type II supernovae.
For SNIa the currently observed rate for ellipticals 
is poorly known and almost nothing is known about 
the SNIa rate at earlier times. 
The SNIa rate is thought to depend on mass exchange between 
binary stars and theory provides no unambiguous way to estimate 
orbital parameters and binary mass ratios.
The yields of heavy metals in SNIa are thought 
to be known reasonably well from theoretical models while
those from SNII are uncertain because of complexities in 
the pre-explosion evolution, uncertain reaction rates, 
and the unknown amount of processed material that falls back on the 
stellar remnant (Woosley \& Weaver 1995;
Gibson, Loewenstein \& Mushotzky 1997).

In view of these difficulties, the relative influence 
of SNII and SNIa in enriching 
elliptical galaxies and rich clusters must be determined 
largely from observations of the metal abundance 
in stars and gas.
The enrichment history of 
individual ellipticals and associated 
groups of galaxies is difficult 
to access since the total energy released in 
all SNII is sufficient to expel a 
significant fraction of metal-rich gas from these systems.
However, rich clusters of galaxies are massive enough to 
have retained all of the material processed in supernovae 
over cosmic time.
The ``closed box'' nature of rich clusters is supported by 
the relative constancy of the ratio 
of baryonic to total mass for cluster masses
$\gta 5 \times 10^{13}$ $M_{\odot}$ (David 1997).
The data assembled by Renzini (1997) indicates that 
metal enrichment by supernovae in rich clusters 
remains confined within the cluster potential;
${\cal M}_g / {\cal L}_B$ and the total iron mass 
per unit optical light 
${\cal M}_{Fe,g} / {\cal L}_B$ are 
nearly constant for cluster masses
$\gta 5 \times 10^{13}$ $M_{\odot}$.
Clusters and galaxy groups with masses less than this 
have lower gas to stellar mass ratios, suggesting 
gaseous outflow.


In determining the past history of supernova enrichment,
the silicon abundance is more useful than the iron abundance.
The silicon abundance has been 
observed in the ICM of many rich clusters 
and is based on a well-understood 
K-line emission feature so that translation into 
abundances should be reliable.
In addition, silicon yields from SNII and possibly SNIa may be 
better determined from theoretical supernova models than 
those of iron.
While the iron yield from SNII $y_{Fe,II}(m)$ increases monotonically 
with the pre-supernova stellar mass $m$, the silicon yield 
$y_{Si,II}(m)$
has a pronounced maximum at $m \sim 22$ $M_{\odot}$ provided 
the energy released by the supernova is not too large
(A and B models of Woosley \& Weaver 1995).
Because of this maximum, the IMF-averaged yield 
$\langle y_{Si,II} \rangle$ is very insensitive to the 
choice of parameters for power law IMFs provided the upper 
mass limit $m_u \gta 22$.

Assuming that silicon is a reliable tracer of past 
supernova activity in rich clusters, 
we wish to construct an expression for the observed 
ratio of the total mass of silicon in the hot cluster gas 
(ICM) to 
the total optical luminosity of spheroidal system 
stars in the cluster,
$$\Upsilon_{Si,g} = {\cal M}_{Si,g} / {\cal L}_B.$$
The present mass of silicon in the ICM is determined 
by the total mass of silicon produced by SNII and SNIa 
less the amount of silicon currently within stars:
\begin{equation}
\Upsilon_{Si,g} \equiv 
{ {\cal M}_{Si,II} + {\cal M}_{Si,Ia} - {\cal M}_{Si,*}
\over  {\cal L}_B }.
\end{equation}
Another important observed quantity is 
the currently observed silicon to iron ratio in the ICM,
\begin{equation}
R \equiv 
\left( {z_{Si} \over z_{Fe}} \right)_g = 
\left( { {\cal M}_{Si} \over {\cal M}_{Fe} } \right)_g
= {
{\cal M}_{Si,II} + {\cal M}_{Si,Ia} - {\cal M}_{Si,*} \over
{\cal M}_{Fe,II} + {\cal M}_{Fe,Ia} - {\cal M}_{Fe,*} }.
\end{equation}
Note that $R$ is the ratio of silicon to iron masses in 
absolute units and is not normalized with the solar ratio.
We now seek expressions for $\Upsilon_{Si,g}$ and $R$ in 
terms of the expected number of supernovae 
and the average silicon and iron yields for each supernova event.

If we assume that all stars more massive than 8
$M_{\odot}$ produce SNII, the number of SNII 
per $M_{\odot}$ of stars formed, $\eta_{II}$, can be found
for any assumed IMF. 
For a power law IMF $\phi(m) dm = \phi_o m^{-(1+x)} dm$ the
number of SNII is equal to the total number
of stars more massive than $m_8 = 8$ $M_{\odot}$, therefore 
$$ \eta_{II} = { N_{II} \over M_*} = {x - 1 \over x}
{ m_8^{-x} - m_u^{-x} \over m_{\ell}^{1-x} - m_u^{1-x}}.$$
For example, 
$\eta_{II} = 6.81 \times 10^{-3}$ SNII per $M_{\odot}$ 
for a Salpeter IMF (slope $x = 1.35$)
having upper and lower masses of $m_u = 100$ $M_{\odot}$
and $m_{\ell} = 0.08$ $M_{\odot}$ respectively.
In recent discussions of the metal enrichment of early type 
galaxies and galaxy clusters it has been fashionable 
to consider only single power law IMFs 
(e.g. Loewenstein \& Mushotzky 1996; Ishimaru \& Arimoto 1997;
Gibson, Loewenstein \& Mushotzky 1997) with 
the Salpeter slope regarded as ``normal.''
However, many lines of evidence in our own Galaxy 
suggest that the normal IMF has some curvature, 
becoming flatter at subsolar masses (Leitherer 1998).
Scalo (1998) suggests the following triple power law 
approximation for a universal IMF: 
$$\phi(m) dm = \phi_i m^{-(1 + x_i)} dm~~~i = 1,2,3$$
with
\begin{eqnarray}
	x_1 = 0.2 \pm 0.3 ~~~{\rm for}~ 0.1 < m < 1~~~ M_{\odot}
		\nonumber\\
        x_2 = 1.7 \pm 0.5 ~~~{\rm for}~ 1.0 < m < 10~~ M_{\odot}
		\nonumber\\
        x_3 = 1.3 \pm 0.5 ~~~{\rm for}~ 10~ < m < 100~ M_{\odot}
		\nonumber
\end{eqnarray}
Evaluating the specific SNII frequency for this IMF 
we find $\eta_{II} = 7.81 \times 10^{-3}$ which differs by less 
than 15 percent from the single-slope Salpeter value. 
In view of the many larger uncertainties involved in 
other parameters, we shall continue to use simple power 
law IMFs in this paper to determine the number of SNII, 
mean Type II supernova yields and the 
stellar mass loss rate.

The total number of SNII explosions 
during the entire history of rich cluster stars 
is ${\cal N}_{II} = \eta_{II} {\cal M}_{*o}$ where 
${\cal M}_{*o}$ is the total initial stellar mass in 
cluster E + S0 galaxies.
The current mass of all early type galaxies in a rich cluster
can be estimated from the total B-band luminosity of all 
cluster galaxies, 
${\cal M}_* = (M/L_B) {\cal L}_B$, where $M/L_B \approx 7$
is an average mass to light ratio for bright 
cluster ellipticals.
Assuming that star formation in ellipticals can be 
approximated with a single burst at time $t_*$,
the stellar mass at that time was 
${\cal M}_{*o} = {\cal M}_* / (1 - \beta)$
where $\beta$ is the fraction of the initial mass that has 
been lost from the stars between $t_*$ and the present 
time $t_n$.
For example, if stars form with a Salpeter IMF at time 
$t_* = 2$ Gyrs and $t_n = 13$ Gyrs is the present time, 
then $\beta = 0.3$. 

Since galactic stars in rich clusters 
are observed to be enriched by SNII ejecta,
it is clear that the single burst approximation is only 
an approximation.
However, if most of the stars are formed in several bursts
concentrated near time $t_*$, the long term stellar evolution will 
be little altered although the single-burst value of $\beta$ will 
be overestimated if some supernova-processed gas 
is formed into stars in nearly simultaneous multi-bursts.
If a fraction $f_*$ of SNII-produced metals formed into stars
at time $t_*$, then a fraction $F_* = f_*(1 - \beta)$ 
is still locked in stars today.
The total mass of SNII-produced silicon 
still remaining in the intracluster gas phase today is therefore
$$(1 - \beta)^{-1} (M/L_B) {\cal L}_B \eta_{II} 
\langle y_{Si,II} \rangle ( 1 - F_*)$$
where the mean silicon yield $\langle y_{Si,II} \rangle$ 
is the IMF-averaged mass of silicon 
(in solar masses) generated per SNII event.
The total mass of iron in the ICM from SNII is given by the same 
expression with $\langle y_{Si,II} \rangle$ replaced 
with $\langle y_{Fe,II} \rangle$. 
By using the single burst assumption, modified by allowing 
some of the SNII ejecta to form stars, we are neglecting 
higher order details such as second generation SNII formed 
from stellar ejecta.


The total amount of silicon produced in SNIa is the 
product of the total number of Type Ia supernovae and 
the silicon yield, ${\cal N}_{Ia} y_{Si,Ia}$.
Since the past rate of SNIa explosions in ellipticals 
is poorly known,
we represent our ignorance with a power law:
$$ { d {\cal N}_{Ia} \over dt} = 
{ {\cal L}_B \over 10^{10} ~ 100 {\rm yrs} }
{\rm SNu}(t) ~~~~ {\rm SNIa~~yr}^{-1} $$
where ${\cal L}_B$ is in units of $L_{B,\odot}$ and 
$$ {\rm SNu}(t) = {\rm SNu}(t_n) (t / t_n)^{-p}$$
is the SNIa rate in SNu units 
(supernovae per $10^{10} L_B$ per 100 years).
The total number of SNIa since $t_*$ is therefore
$${\cal N}_{Ia} = {\cal L}_B { {\rm SNu}(t_n) \over 10^{10}~100}
t_n \int_{t_*/t_n}^1 (t/t_n)^{-p} d(t/t_n) \equiv {\cal L}_B n_{Ia}$$
where $n_{Ia}$ is the number of SNIa per unit $L_B$.
The important parameter that governs 
the SNIa contribution to the observed ratios $R$ and 
$\Upsilon_{Si,g}$ is $n_{Ia}$ although we shall occasionally 
use the current SNIa rate SNu$(t_n)$ to characterize 
$n_{Ia}$, assuming particular values of $p$ and $t_*$. 
If a fraction $g_*$ of the enriched SNIa ejecta 
was incorporated into stars at time $t_*$, then a fraction 
$G_* = g_*(1 - \beta)$ remains in stars at the present time.

Finally, the current stellar iron abundance is given by 
\begin{equation}
z_{Fe,*} = 1.4 { {\cal M}_{Fe,*} \over {\cal M}_* }
\end{equation}
where $z_{Fe,*}$ is the abundance of iron by mass in stars
relative to hydrogen and $z_{Fe,*}/1.4$
is the ratio of iron mass to total mass including helium.

Following the notation introduced above,
Equations (1) - (3) can be written as 
\begin{equation}
\Upsilon_{Si,g} = { {\cal M}_{Si,g} \over {\cal L}_B }
= (1 - \beta)^{-1} (M/L_B) \eta_{II} 
\langle y_{Si,II} \rangle (1 - F_*) 
+ n_{Ia} y_{Si,Ia} (1 - G_*),
\end{equation}
\begin{equation}
R = \left( { {\cal M}_{Si} \over {\cal M}_{Fe} } \right)_g
= {
(1 - \beta)^{-1} (M/L_B) \eta_{II} \langle y_{Si,II} \rangle 
(1 - F_*)
+ n_{Ia} y_{Si,Ia} ( 1 - G_*)
\over
(1 - \beta)^{-1} (M/L_B) \eta_{II} \langle y_{Fe,II} \rangle 
(1 - F_*)
+ n_{Ia} y_{Fe,Ia} ( 1 - G_*)
}.
\end{equation}
and
\begin{equation}
z_{Fe,*} = 1.4[ (1 - \beta)^{-1} \eta_{II} 
\langle y_{Fe,II} \rangle F_* 
+ (M / L_B)^{-1} n_{Ia} y_{Fe,Ia} G_* ].
\end{equation}
If SNIa do not contribute to the enrichment of the stars,
then the ratio of stellar silicon and iron must be in
proportion to SNII yields, $z_{Fe,*} = z_{Si,*}
\langle y_{Fe,II} \rangle / \langle y_{Si,II} \rangle$;
equation (6) reduces to this simple expression when 
$n_{Ia} = 0$ or $G_* = 0$.
With the total number of SNIa 
regarded as a parameter,
we shall solve the three equations above for the most 
uncertain remaining parameters:
the specific number of SNII, $\eta_{II}$,
the mean iron yield from SNII $\langle y_{Fe,II} \rangle$,
and the fraction $f_*$ of SNII ejecta that has formed into
stars.

Values for the many additional parameters 
in the last three equations must be determined from 
theoretical supernova calculations or from observation.
IMF-averaged theoretical supernova yields for iron and silicon have been 
collected and discussed by Loewenstein \& Mushotzky (1996)
and Gibson, Loewenstein \& Mushotzky (1997).
Following these authors we take 
SNIa silicon and iron yields from Model W7 of 
Thielemann, Nomoto \& Hashimoto (1993):
$y_{Si,Ia} = 0.158$ and 
$y_{Fe,Ia} = 0.744$, both in $M_{\odot}$.
Yields for SNII collected by Loewenstein \& Mushotzky (1996), 
mostly based on Woosley \& Weaver (1995),
have a large range for different models:
$\langle y_{Si,II} \rangle = 0.08 - 0.32$ and
$\langle y_{Fe,II} \rangle = 0.11 - 0.34 $ while 
yields for the less extreme SNII models discussed by 
Gibson, Loewenstein \& Mushotzky (1997) 
have a more limited range: 
$\langle y_{Si,II} \rangle = 0.104 - 0.143$ and
$\langle y_{Fe,II} \rangle = 0.073 - 0.141$ 
both evaluated with a Salpeter IMF. 
Gibson, Loewenstein \& Mushotzky (1997)
discuss the uncertainties in theoretical SNII yields in
detail. 
As discussed previously, because of 
the maximum in the silicon yield near progenitor mass 
$22$ $M_{\odot}$, the mean silicon yield 
$\langle y_{Si,II} \rangle \approx 0.133$ $M_{\odot}$ 
appears to be more securely 
known than $\langle y_{Fe,II} \rangle$ which we regard 
as an unknown to be determined by solving the equations above.

The stellar mass to light ratio 
in early type galaxies is a slowly increasing function of 
galactic luminosity, $M / L_B \propto {L_B}^{0.2}$
(Faber et al. 1984), and 
depends somewhat on the cluster luminosity function. 
For a representative value 
we choose $M / L_B = 7$ (in solar units),
characteristic of typical bright ellipticals.

Values of $\Upsilon_{Si,g}$ and $R$ in Equations (4) and (5) 
must be supplied from X-ray observations. 
For three rich clusters (A 2199, A 496, and AWM 7)
observed with ASCA 
Mushotzky et al. (1996)
find an average silicon mass to light ratio 
$\Upsilon_{Si,g} = 0.0305 \pm 0.008$ in solar units. 
More extensive observational 
data indicates that there is a significant 
cosmic variation in the abundances of Si and Fe
among rich clusters (Mushotzky \& Loewenstein 1997;
Mushotzky 1998).
If real, this variation of galaxy-averaged values suggests that 
the IMF, star formation efficiency, supernova frequency or other 
supernova properties may vary dramatically 
among cluster galaxies.
Nevertheless, for our purposes here we simply use 
average abundances from all of the clusters observed.
ASCA observations indicate a silicon to iron ratio of 
Si/Fe = $2.2 \pm 0.25$ (Mushotzky 1998) in solar units.
But this abundance ratio, 
based on the so-called ``photospheric'' solar iron abundance, 
becomes Si/Fe $= 2.2(1.83/2.66) = 1.51$ in units of 
the ``meteoritic'' solar ratio (Si/Fe)$_{\odot} = 0.550$ 
adopted here. 
The corresponding absolute Si/Fe abundance ratio 
in equation (5) is therefore $R = 0.83$

Solutions of Equations (4) - (6) 
for $\eta_{II}$, $\langle y_{Fe,II} \rangle$, 
and $f_*$ are shown in Figure 1 as functions of 
the total number of SNIa events ${\cal N}_{Ia}$ 
and the current Type Ia supernova rate SNu$(t_n)$
evaluated with $p = 1$, $t_n = 13$ Gyrs, and $t_* = 2$ Gyrs.
Two representative 
solutions are illustrated: (i) with no enrichment of stars 
by SNIa ejecta $g_* = 0$ and (ii) with a 
significant SNIa enrichment $g_* = 0.5$.
The plotted solutions are based on a mass return parameter
appropriate for a single burst Salpeter IMF $\beta = 0.3$. 

Type Ia supernova rates observed in elliptical galaxies
favor low values of SNu$(t_n)$ where 
$\eta_{II}$, $f_*$ and $\langle y_{Fe,II} \rangle$ in 
Figure 1 are almost independent of SNu$(t_n)$
and ${\cal N}_{Ia} \ll {\cal N}_{II}$.
In a recent study of the observed frequency of SNIa in
ellipticals,
Cappellaro et al. (1997) find a low current SNIa rate
SNu$(t_n) = 0.058 (H/50)^2$ in SNu units.
The current SNIa rate is also restricted by 
our gas dynamical models for the evolution of hot gas
in large elliptical galaxies. 
The computed radial variation of gas 
density and temperature agree with those observed only when 
SNu$(t_n) \lta 0.25$ (with $p = 1$), 
again suggesting low values for ${\cal N}_{Ia}$.
However, when silicon and iron abundances are included 
in these hydrodynamical evolution, as discussed below,
low values of SNu$(t_n)$ and ${\cal N}_{Ia}$ are essential.
In this low-SNu$(t_n)$ range of Figure 1, solutions for 
the three parameters
$\eta_{II}$, $\langle y_{Fe,II} \rangle$,
and $f_*$ approach the limit of no SNIa enrichment, 
${\cal N}_{Ia} \rightarrow 0$.
In this limit solutions to Equations (4) - (6) for 
$\Upsilon_{Si,g}$, $R$, and $z_{*Fe}$ simplify to
\begin{equation}
\langle y_{Fe,II} \rangle = \langle y_{Si,II} \rangle / R
= 0.16,
\end{equation}
\begin{equation}
\eta_{II} = { \Upsilon_{Si,g} + R (z_{Fe,*}/1.4) (M/L_B)
\over (M / L_B) \langle y_{Si,II} \rangle (1 - \beta)^{-1} }
= 0.026,
\end{equation}
and
\begin{equation}
f_* = {R (z_{Fe,*}/1.4) \over \eta_{II} \langle y_{Si,II} \rangle }
= 0.16.
\end{equation}
These numerical evaluations are based on
$\Upsilon_{Si,g} = 0.0305$, $R = 0.83$,
$z_{Fe,*} = 0.5 z_{Fe,\odot}$, $M/L_B = 7$,
$\langle y_{Si,II} \rangle = 0.133$, and $\beta = 0.3$.
Since $\Upsilon_{Si,g}$ dominates the numerator of Equation (8),
the value of $\eta_{II}$ is essentially unchanged if the 
mean stellar iron abundance is taken to be solar, 
$z_{Fe,*} = z_{Fe,\odot}$.

The value of $\langle y_{Fe,II} \rangle$ in Equation (7)
is within the range of possible SNII iron yields calculated
by Woosley \& Weaver (1995); this provides some confidence 
in the veracity of other parameters determined 
from the equations above. 
However, the specific supernova rate
$\eta_{II} = 0.026$ is almost four times the value
expected for a Salpeter IMF, $\eta_{std} = 0.00681$
($x = 1.35$, $m_u = 100$ $M_{\odot}$
and $m_{\ell} = 0.08$ $M_{\odot}$).
Values of $\eta_{II}$ computed with single power law IMFs are 
sensitive to both the slope $x$ of the IMF and the
mass limits $m_u$ and $m_{\ell}$.
The variation of
$\eta_{II}$ and $\beta$ with $x$, $m_u$ and $m_{\ell}$
for several power law IMFs is illustrated in Figure 2.

It is difficult to assign errors to the quantities 
evaluated in Equations (7) - (9) because of the many uncertainties
involved and the possibility of systematic errors.
If all the quantities in Equation (8) are skewed to their
limits of uncertainty in an effort to minimize $\eta_{II}$, it
is possible to achieve a close agreement with a Salpeter IMF.
For example, if 
$\Upsilon_{Si,g} = 0.0225$, $R = 0.92$,
$M/L_B = 9$,
$\langle y_{Si,II} \rangle = 0.32$ (model WW1ex from Loewenstein
\& Mushotzky 1996), and $\beta = 0.3$
then $\eta_{II} = 0.0068$, close to the Salpeter value.
Of course it is most unlikely 
that Nature would conspire in this manner.

Also shown in Figure 1 are the fraction of all iron  
created by SNIa:
$$F_{Ia} = { n_{Ia} y_{Fe,Ia} 
\over
n_{Ia} y_{Fe,Ia}
+ (1 - \beta)^{-1} (M/L_B) \eta_{II} 
\langle y_{Fe,II} \rangle }$$
and the fraction of all iron {\it in the hot ICM gas} 
that originated in SNIa:
$$F_{Ia,g} = { n_{Ia} y_{Fe,Ia} (1 - G_*)
\over
n_{Ia} y_{Fe,Ia} (1 - G_*)
+ (1 - \beta)^{-1} (M/L_B) \eta_{II}
\langle y_{Fe,II} \rangle (1 - F_*)  }.$$
Provided SNu$(t_n)$ is limited to values of interest, 
$\log[{\rm SNu}(t_n)] \lta -1$, 
Type Ia supernovae can contribute to the stellar 
enrichment without substantially changing 
$\eta_{II}$ or $\langle y_{Fe,II} \rangle$.

Renzini et al. (1993), Ishimaru \& Arimoto (1997), 
Renzini (1997) and Wyse (1997) 
have discussed the X-ray data for rich clusters in detail.
These authors prefer models in which 
the overabundance of iron relative to that 
expected with a normal Salpeter IMF 
is due to a large additional 
iron contribution from Type Ia supernovae,
corresponding to $F_{Ia} \sim 0.5$ in Figure 1.
Such a model would be very similar to the enrichment history of 
our own Galaxy.
In particular, Renzini et al. (1993) suggest a model for 
rich cluster enrichment in which $\sim {3 \over 4}$ of 
the iron is produced by SNIa 
with the remaining $\sim {1 \over 4}$ coming from SNII.
If only half of the iron in rich clusters has originated in SNIa, 
then the number of SNIa per $L_B$ is 
$n_{Ia} = {\cal N}_{Ia}/{\cal L}_B \approx 0.025$.
But such a large number of SNIa is incompatible with 
iron and silicon abundances in 
our calculated model of a single large elliptical, 
discussed below, where we 
find $n_{Ia} = {\cal N}_{Ia}/{\cal L}_B \lta 0.001$. 
For the simple power law model for SNu$(t)$ we 
adopt in Figure 1 
($p = 1$, $t_* = 2$ Gyrs, $t_n = 13$ Gyrs), 
the current SNIa rate would need to be very 
large, SNu$(t_n) \gta 0.6$, for Type Ia supernova to 
supply half of the iron.
This is about ten times greater than the SNIa rate 
estimated from observations of bright ellipticals.

An alternative explanation of the iron excess 
in rich clusters relative that produced by a Salpeter IMF
($x$, $m_u$, $m_{\ell}$ = 1.35, 100, 0.08) 
is to invoke a flatter IMF having a larger fraction 
of massive stars and associated SNII.
This is the 
interpretation preferred by
David (1997) and Gibson, Loewenstein \& Mushotzky (1997).
For consistency both $\beta$ and $\eta_{II}$ must 
be adjusted when the IMF slope is changed; 
the value of the mass return fraction $\beta = 0.3$
used in Equation (8) is based on the 
Salpeter IMF. 
Logically compatible parameters can be found 
by simultaneously solving 
Equation (8) for $\eta_{II} ( \beta)$
with the parametric
variations $\eta_{II}(x)$ and $\beta(x)$ plotted in Figure 2.
The result of this joint solution is that Equation (8)
now gives $\eta_{II} = 0.016$ with $\beta = 0.55$
for ($x$, $m_u$, $m_{\ell}$ = 1.00, 100, 0.08)
or $\eta_{II} = 0.018$ with $\beta = 0.51$
for ($x$, $m_u$, $m_{\ell}$ = 0.94, 40, 0.08).
(Such large values of the mass return $\beta$ 
from evolving stars
may be unrealistic since several generations of
stars must have formed near time $t_*$, 
each consuming some of the metal-enriched 
gas ejected from previous generations of stars.)
With these flat IMFs 
almost all of the iron in rich clusters originates in SNII 
and the low SNIa rates are consistent with
those observed in massive ellipticals.
IMFs flatter than Salpeter 
are also supported by the large mass of oxygen 
in rich clusters; 
Gibson, Loewenstein \& Mushotzky (1997) claim that the 
observed O/Si ratio can be produced by SNII alone with 
little or no contribution from SNIa.

When silicon and iron abundances in the ICM  
are considered together with abundance constraints 
set by individual massive ellipticals, as we have done here, 
the metal enrichment of rich clusters 
by early type galaxies is shown to be 
almost exclusively due 
to Type II supernovae with a negligible contribution from 
Type Ia supernovae. 
Evidently, 
star formation conditions in rich clusters are quite unlike 
those that prevail in our own Galaxy.

The controversy concerning the relative enrichment from 
Type II and Type Ia supernovae in rich clusters 
is further confounded by the  
gas dynamical solutions we discuss below.
We describe a variety of supernova enrichment 
histories for massive elliptical galaxies 
in which the hot interstellar gas is constrained 
to evolve toward the radial 
distribution of gas density, temperature and metal abundance 
observed today.
In particular 
we show that the production of SNII using a 
Salpeter IMF is sufficient to achieve simultaneous 
agreement with $n(r)$, $T(r)$ and $z_{Fe}(r)$ 
indicated by X-ray observations of massive ellipticals. 
Gas dynamic solutions based on 
IMFs flatter than Salpeter are generally incompatible with 
the observed abundance variation $z_{Fe}(r)$.
The SNIa rate must also be low.
If more supernovae of either type are involved, 
as indicated by the large 
$\eta_{II}$ for rich clusters (Equation 8), 
the iron and silicon abundances
in the models greatly exceed those observed in 
massive elliptical galaxies.


\section{IRON AND SILICON IN NGC 4472}

\subsection{Gas Dynamical Models for NGC 4472}

In a recent paper (Brighenti \& Mathews 1998b) we discuss 
in detail the basic assumptions and equations used 
in our gas dynamical models that simulate the evolution 
of hot interstellar gas in elliptical galaxies.
Since the models we discuss here are 
very similar to those in that paper, 
we provide only a brief summary.

Our 1D spherical calculations begin with 
an overdensity perturbation in a flat cosmology 
having an amplitude designed to produce a mass concentration 
similar to that of NGC 4472 after a few gigayears.
The flow of dark and baryonic matter far from the 
center of the perturbation follows
the self-similar solution described by Bertschinger (1985).
An outward moving 
turn-around radius defines the instantaneous 
locus where the cosmic flow velocity vanishes.
The ``secondary infall'' occurs within the turn-around radius, 
allowing baryonic and dark matter to collect near the origin.
The central 
accumulation of dark matter grows from the inside out;
although the collisionless dark fluid interpenetrates, 
after passing through the origin it continues to oscillate,
spending most of its time at large radii.
The net effect is that a quasi-stationary core of dark matter 
grows outward from the origin 
and is matched to the outer Bertschinger inflow in a manner that 
preserves the total mass of dark matter.
However, we replace the inner power law core of the 
self-similar Bertschinger solution 
with a (less peaked) 
dark halo having an NFW profile as determined by 
Navarro et al. (1996) with full three-dimensional
N-body calculations. 
The dark matter mass of NGC 4472 inferred by X-ray data 
can be fit reasonably well
with an NFW dark halo of mass 
$M_h = 4 \times 10^{13}$ $M_{\odot}$ within the current 
virial radius assuming $D = 17$ Mpc for the distance to NGC 4472.
The cold baryonic gas also participates in the Bertschinger flow, 
but deep within the turnaround radius 
it passes through an accretion shock, compresses and is 
heated to approximately the virial temperature of the galactic 
dark halo.
At time $t_*$ when enough baryonic matter has collected within the shock,
some of which has radiatively cooled, we form the 
de Vaucouleurs stellar configuration having a total mass 
$M_* = 7.26 \times 10^{11}$ $M_{\odot}$ 
appropriate to NGC 4472 and corresponding 
to a stellar mass to light ratio 
$M_{*}/L_B = 9.20$ (van der Marel 1991).
The stellar galaxy is constructed by removing a baryonic mass 
equal to $M_*$ from gas within the 
accretion shock $r_{sh}(t_*)$ to form the stars. 
The gas density in the remaining gas is 
reduced in proportion to its density just before $t_*$.

We assume the SNII energy is released immediately at time $t_*$.
After gas has been removed to form the stars,
the SNII energy is assumed to be evenly deposited 
(masswise) to remaining gas within the 
accretion radius or within some other specified radius.
Removing gas interior to $r_{sh}(t_*)$ and heating 
the remaining gas introduces a transient in the flow,
but after $\sim 1$ Gyr 
gas moving within $r_{sh}$ becomes subsonic and 
approaches hydrostatic equilibrium.
The subsequent time dependent 
gas dynamics within and around the galaxy 
are followed in full detail.
The potential of the dark matter continues to evolve 
according to the NFW-Bertschinger prescription and the 
baryonic shock grows in radius.
The flow of these two fluids 
is solved simultaneously using 1D Eulerian hydrodynamics 
on a logarithmic grid.
The equations that describe this flow 
are described in detail in our 
recent papers (e.g. Brighenti \& Mathews 1998b).
Inside the de Vaucouleurs core 
new gas is provided by mass loss from 
stars as they evolve off the main sequence 
according to some assumed stellar IMF.
Most of the gas within the accretion shock radiates thermal
X-rays corresponding to a temperature $T \sim 10^7$ K 
similar to the virial temperature in the total potential.
After time $t_*$ additional energy is supplied to the gas 
by the dissipation of the orbital energy of mass-losing 
stars and by Type Ia supernovae.
The stellar velocity field is found from solutions of the 
Jeans equation in the combined stellar and dark matter 
potential, 
although the evolved solutions are (fortunately) 
rather insensitive to the precise variation of the stellar 
velocity ellipsoid with galactic radius, as characterized 
by the stellar temperature $T_*(r)$.
This galaxy formation model is obviously oversimplified, 
but it properly conserves both dark and baryonic matter 
and provides a realistic template for our gas dynamical 
models.

In a typical calculation the shock radius at time 
$t_*$ is $r_{sh}(t_*) \sim 200$ kpc, considerably larger 
than the present size of the stellar system for NGC 4472
(effective radius $r_{e} = 8.57$ kpc, 
maximum stellar radius $r_{*t} = 100$ kpc).
It may seem inconsistent to introduce SNII heating and 
metal enrichment within a radius $r_{sh}(t_*)$ 
that is much larger 
than the currently observed size of the stellar galaxy.
This apparent inconsistency can be 
justified because 
star formation must have occurred in a somewhat larger 
region than is occupied by old stars in NGC 4472 today;
in order to form the de Vaucouleurs profile in NGC 4472
by violent relaxation, it is necessary that the pre-galactic 
objects be largely stellar before the galaxy is assembled.

The total energy released by SNII explosions at time $t_*$ is 
$E_{II} = \epsilon_{sn} \eta_{II} M_* E_{sn}$ 
where $E_{sn} = 10^{51}$
ergs, ${\cal N}_{II} = \eta_{II} M_*$ 
is the total number of SNII
produced and $\epsilon_{sn}$ is the efficiency that SNII 
energy is converted to thermal energy in the ambient gas,
the rest is lost to radiation.
For a reference value of $\eta_{II}$ we use the value 
computed with a Salpeter IMF 
($x$, $m_u$, $m_{\ell}$ = 1.35, 100, 0.08),
$\eta_{std} = 6.81 \times 10^{-3}$.
For simplicity, the stellar mass in these relations 
(and in establishing the fixed stellar potential) 
is set to the current 
value $M_* = M_*(t_n)$ rather than its value at $t_*$,
$M_*/(1 - \beta)$. 
In view of the uncertainties involved in all these parameters, 
in the gas dynamical models we 
evaluate $E_{II} = \eta_{II} M_*(t_n) 10^{51}$ 
as if $\epsilon_{sn} = 1$ 
and vary $\eta_{II}$, seeking results that best match 
current observations.
Each $E_{II}$ determined in this way corresponds to 
a range of 
$M_*(t_*) E_{sn}$ for each assumed 
gas heating efficiency 
$\epsilon_{sn} = E_{II}/ \eta_{II}  M_*(t_*) E_{sn} \leq 1$. 
Metal enrichment by SNII accompanies the energy deposition 
within the accretion shock radius at $t_*$.
Altogether 
$\langle y_{Fe,II} \rangle \eta_{II} M_*$ solar masses of iron 
are available for enriching stars and gas within $r_{sh}(t_*)$.
If $z_{Fe,*}$ is the assumed stellar iron abundance, a mass
$z_{Fe,*} M_*/1.4$ of iron produced by SNII is allocated to the 
newly formed stellar system at time 
$t_*$ and the remainder is introduced 
into the gas within the shock radius in proportion to the
local gas density.

In summary, the important galactic 
parameters that determine the early star formation
and SNII enrichment in rich clusters and individual 
massive ellipticals 
are (i) the total energy released in SNII explosions 
$E_{II}$, 
(ii) the number of SNII per solar mass $\eta_{II}$, and
(iii) the time of star formation $t_*$.

\subsection{The Standard Model}

In our standard or reference 
model for the evolution of hot interstellar gas 
in NGC 4472, the galaxy forms in a flat universe
($\Omega = 1$) with global baryon fraction $\Omega_b = 0.05$.
The specific SNII rate is 
$\eta_{II} = \eta_{std} \equiv 6.81 \times 10^{-3}$.
The specific stellar mass loss rate for the Salpeter IMF 
$\alpha_*(t) = d \log M_* / dt$ is well fit with 
a power law
$\alpha_*(t) = 4.7 \times 10^{-20}
(t/t_s)^{-1.26}$ s$^{-1}$
where $t_n = 13$ Gyrs and and $t_s = t_n - t_*$
is the current age of the stars.
Both stars and SNII are assumed to form at time 
$t_* = 2$ Gyrs.
The total number of SNIa, ${\cal N}_{Ia}$, 
is an integral over SNu$(t)$ which is assumed to 
vary as a simple power law 
parameterized with SNu$(t_n) = 0.015$ SNu and 
$p = 1$. 
Type Ia supernovae are assumed to begin at $t_*$ and 
continue until $t_n = 13$ Gyrs. 
Except for the slightly lower SNu$(t_n)$, 
all parameters are identical to the standard reference 
model discussed in Brighenti \& Mathews (1998b).
In addition, we assume for the standard model that no
iron or silicon produced by SNIa is used in stellar
enrichment, so the stars have abundances proportional
to SNII yields.

The first requirement for an acceptable model 
is that the chosen parameters, 
when used in a gas dynamical calculation, 
adequately reproduce the radial variation of interstellar
gas density and temperature observed in NGC 4472 today.
In Figure 3 we plot with solid lines 
the density and temperature profiles
in the hot interstellar gas for the standard 
model at time $t_n = 13$ Gyrs.
The overall agreement with the observed gas temperature 
and density is satisfactory but not perfect.
The excess gas density in the model 
in $r \lta r_e = 8.57$ kpc is a classical artifact 
of galactic cooling flow models; we believe that it 
can be mitigated by galactic rotation 
and low mass star formation but this has not 
yet been adequately demonstrated (see 
Mathews \& Brighenti 1998c for a brief review).
At larger radii, $r \gta 10$ kpc, the computed density
is slightly low.
This latter discrepancy may be due to the disturbed 
nature of NGC 4472 at large radii where it appears 
to be interacting with ambient gas in the Virgo cluster
(Irwin \& Sarazin 1996) although the 
azimuth-averaged outer gas density profile in 
NGC 4472 is typical of the X-ray structure 
in other large ellipticals (Mathews \& Brighenti 1998b).
In any case, 
if $E_{II}$ is slightly lower than the Salpeter 
value, the standard model 
can be adjusted to fit the observations 
almost perfectly at large $r$ (Brighenti \& Mathews 1998a).

For the models discussed here we assume a simple
flat universe, $\Omega = 1$, but it is also possible 
to find satisfactory and very similar 
dynamical models for NGC 4472 using other cosmologies.
In the Appendix we briefly describe 
gas dynamical models appropriate for an open
universe ($\Omega = 0.3$) and a low density flat
universe ($\Omega = 0.3$, $\Omega_{\Lambda} = 0.7$).

A fully successful model must also match the 
metal abundances and abundance gradients 
currently observed in NGC 4472.
Unfortunately, supernova yields and observed 
abundances are both uncertain so there is some flexibility 
in making this fit.
The iron abundance in NGC 4472 
has been observed extensively with ROSAT and 
ASCA by many observers using many different 
data reduction procedures.
Observed iron abundances expressed in solar units 
can be a source of some confusion if authors 
do not explicitly note which iron abundance they 
have assumed for the sun, photospheric or meteoritic.
If the lower 
meteoritic value is more correct, as we assume here,
iron abundances relative to the solar photospheric value 
are too low by 1.44 (Anders \& Grevesse 1989).
In the following discussion, we consider only 
observational sources 
for which the absolute solar abundance is specified and, 
if necessary, we convert observed iron abundances  
cited in the literature to meteoritic solar.
Unfortunately, observational determinations of the iron 
abundance in the hot gas also seems to depend on the 
data reduction procedure used.
Some of the difficulty in deriving hot gas 
abundances arises due to the presence of additional 
X-ray emission unrelated to the hot interstellar gas.
Harder radiation ($E \gta 2$ keV) thought to be stellar in 
origin must be allowed for in fitting the spectrum 
and deriving abundances. 
Observational determinations of the iron abundance 
appear to depend on 
the procedure used in allowing for the hard radiation.
For example, 
Buote and Fabian (1998) find that the iron abundance in 
NGC 4472 determined with a single temperature plasma
increases by a factor of 4.7 when a two-temperature
fit is used instead.
Possible 
inadequacies in treating the Fe L transitions 
in the adopted plasma code may also influence the resulting 
value for the observed iron abundance.

Seeking some consistency, we consider here 
several published globally averaged iron 
abundances determined with detectors on the ASCA satellite. 
Using the Raymond-Smith 
plasma code with a hard component in the spectrum, 
Arimoto et al. (1997) find $z_{Fe} = 0.47 \pm 0.09$ solar 
in the hot interstellar gas in NGC 4472 within 3 arc minutes
(15 kpc) of the galaxy center.
Matusmoto et al. (1997) use a fixed hard spectrum 
typical of low mass binary stars and find 
$z_{Fe} = 0.56 \pm 0.09$ solar for NGC 4472.
Finally, 
Matsushita (1997) gives $z_{Fe} = 0.63 \pm 0.08$ solar 
for the iron abundance in NGC 4472.
All these values are consistent at about half solar, 
but Buote \& Fabian (1998), also using ASCA data 
with the MEKAL code,
derive two quite different abundances: 
$z_{Fe} = 0.37$ solar 
(single temperature model: T = 0.97 keV) and 
$z_{Fe} = 1.70 \pm 0.46$ solar 
(two temperature model: 
$T_1 = 0.76$ keV, $T_2 = 1.48$ keV).
All these values are emission-weighted ($\propto \rho^2$) 
and therefore apply mostly to denser interstellar gas 
located closer to the galactic center.
In addition to these globally averaged values,
Matsushita (1997) finds a negative iron abundance gradient 
in several bright ellipticals including NGC 4472; 
this gradient will play an important role in fitting 
to abundances computed in our gas dynamical models.
Because of the additional spatial information provided, 
we adopt Matsushita's iron abundances for comparison
with our models. 
Finally, Matsushita (1997) and Arimoto et al. (1997) find a global
silicon abundance $z_{Si} \approx 0.5$ solar 
for NGC 4472.

Figure 4 illustrates the distribution of iron and silicon 
in the hot gas surrounding NGC 4472 after evolving to 
the present time $t_n = 13$ Gyrs.
This calculation is based on 
supernova parameters for the standard 
model --  SNu$(t_n)$ = 0.015, $p = 1$ and 
$\eta_{II}=6.81 \times 10^{-3}$ -- with  
supernova yields: $\langle y_{Fe,II} \rangle=
0.14$ M$_\odot$, 
$\langle y_{Si,II} \rangle=0.133$  M$_\odot$, $y_{Fe,Ia}=0.744$
M$_\odot$, and $y_{Si,Ia}=0.158$ M$_\odot$.
Following Arimoto et al. (1997), 
we assume a power law gradient for the 
stellar metal abundance 
$$ z_{Fe,*} = 0.675 (r / r_e)^{-0.207}$$
which is based on the observed spatial variation of the 
Mg$_2$ photometric index. 
With this variation the mean stellar Fe/Mg ratio is 
half solar, 
$\langle z_{Fe,*} \rangle / \langle z_{Mg,*} \rangle = 0.5$, 
(in solar units) as suggested by Trager (1997)  
where $ \langle z_{Mg,*} \rangle = 1.385$ solar
(Arimoto et al. 1997).
Since SNIa do not enrich the stars in the standard solution, 
the stellar silicon abundance is proportional to SNII yields,
$z_{Si,*} = z_{Fe,*} 
\langle y_{Si,II} \rangle/\langle y_{Fe,II} \rangle$.

In the standard model 
all SNII occur as the 
stellar system is created at time $t_* = 2$ Gyrs 
and the stars are enriched with enough iron and silicon 
from SNII to match the mean stellar abundances 
$\langle z_{Fe,*} \rangle$ and $\langle z_{Si,*} \rangle$.
All additional iron and silicon from SNII 
is distributed at time $t_*$ to gas within the 
accretion shock radius $r_{sh}(t_*) = 200$ kpc in proportion 
to the local gas density.
The energy released by SNII is sufficient to temporarily 
reverse the secondary inflow, causing the contact discontinuity 
that defines the enriched region within $r_{sh}(t_*)$ 
to expand to 400 kpc at the present time 
as shown in Figure 4.
The uniformity of $z_{Fe}$ and $z_{Si}$ in the ``plateau'' region 
visible in Figure 4 between the outermost stars 
and the contact discontinuity, 
100 - 400 kpc, is an artifact of 
the uniform deposition of SNII metals at time $t_*$.
Most of the iron mass is in this plateau region as shown 
in the plot of $M_{Fe}(r)$ in Figure 5.
At time $t_n$ the interstellar gas 
is slowly flowing inward within the galaxy 
$r \lta 100$ kpc,
where the early SNII enrichment is 
diluted by new gas contributed by stellar mass loss. 
But this dilution is more than compensated by additional iron 
and silicon contributed by SNIa and stellar mass loss within the 
galaxy.
The total iron abundance distribution from all of these 
sources, shown with a solid line in Figure 4, 
is seen to agree reasonably well with 
the observations of Matsushita (1997).
Emission-weighted abundances based on this standard model depend 
on the radius considered.
For the entire region within the current accretion shock 
radius $r_{sh}(t_n) = 1000$ kpc the mean abundances in 
the gas (in solar units) are 
$\langle z_{Fe} \rangle = 0.45$
and $\langle z_{Si} \rangle = 0.52$; 
within the current contact discontinuity at 400 kpc 
the mean gas abundances are  
$\langle z_{Fe} \rangle = 0.67$ 
and $\langle z_{Si} \rangle = 0.77$;
within 150 kpc the abundances are even higher,
$\langle z_{Fe} \rangle = 0.76$ and $\langle z_{Si} \rangle = 0.86$.
These latter values are slightly higher than the global 
iron and silicon 
abundances reported by Matsushita (1997); this 
excess probably arises because the gas density in our models is 
slightly larger and 
more centrally peaked than that observed in NGC 4472 
(Figure 3). 
The approximate constancy of $(z_{Si}/z_{Si,\odot})
/(z_{Fe}/z_{Fe,\odot}) \approx 0.6$ 
with galactic radius 
and its value intermediate between SNII and SNIa yields 
($\langle y_{Si,II} \rangle/\langle y_{Fe,II} \rangle 
\approx 0.8$ and 
$y_{Si,Ia}/y_{Fe,Ia} \approx 0.2$ respectively) are 
both attributes 
of recent X-ray observations of NGC 5846 by 
Finoguenov et al (1998).

For any given gas dynamical solution described 
by $n(r)$ and $T(r)$ at time $t_n$,
an infinite family of interstellar abundance distributions 
can be generated by changing the specific number of 
SNII $\eta_{II}$ (or the supernova yields)
while keeping the total SNII energy 
$E_{II} = \epsilon_{sn} \eta_{II} M_* E_{sn}$ unchanged.
The constancy of $E_{II}$ is ensured by 
varying the gas heating efficiency factors 
$\epsilon_{sn}$ or $E_{sn}$ to compensate 
for changes in $\eta_{II}$. 
When following this procedure, 
we were surprised to discover that the iron abundance 
distribution shown in Figure 4 disagrees 
significantly with Matsushita's observations when 
$\eta_{II}$ is only varied by $\sim 10$ percent from that 
used in the standard solution.
For example, if $\eta_{II} = 1.25 \eta_{std}$ 
the iron abundance in the plateau region (100 - 400 kpc) 
increases to $z_{Fe}/z_{Fe,\odot} = 0.72$ 
and the central value is nearly 1.1.
These iron abundances are significantly higher than Matsushita's 
observed values, particularly in the outer galaxy 
$r \gta 20$ kpc.
The sensitivity of $z_{Fe}(r,t_n)$ 
to $\eta_{II}$ can be understood 
because of the large amount of iron and silicon required
to enrich the galactic stars.
The global iron abundance in the gas just after SNII 
enrichment, identical to that in the plateau region,  is 
$z_{Fe} = (M_{Fe,II} - M_{Fe,*})/M_{gas}$
where $M_{Fe,II}$ is the total iron produced by SNII.
Since the large total mass of iron in stars 
$M_{Fe,*}$ is held 
fixed as $\eta_{II}$ is varied to create a 
family of dynamically identical enrichment models, 
small changes in $\eta_{II}$ and therefore 
$M_{Fe,II}$ correspond to rather large changes if $z_{Fe}$ 
since $M_{Fe,II}$ is not much larger than $M_{Fe,*}$.
In this sense our models require a high degree of 
regularity in $\eta_{II}$ in order to match the 
negative gaseous iron abundance gradients typically 
observed in giant ellipticals.
This sensitivity to $\eta_{II}$ 
may be an indication that our model is too simple or that 
some additional enrichment process has been overlooked.

Although our standard model,
as illustrated in Figures 3 and 4,
is successful in 
approximately reproducing all relevant observations of NGC 4472, 
the value of $\eta_{II}$ that we have used corresponds to 
a normal Salpeter IMF 
and is therefore several times less than values 
required to enrich massive clusters of galaxies (Equation 8).
In spite of this important difference in $\eta_{II}$, 
the gaseous iron and silicon abundances in our standard model 
within the current accretion shock radius 
$r_{sh}(t_n) = 1000$ kpc, 
$\langle z_{Fe} \rangle = 0.45~z_{Fe,\odot}$ 
and $\langle z_{Si} \rangle = 0.52~z_{Fe,\odot}$,
are both rather similar to typical
abundances in the ICM of rich clusters, $\sim 0.4$ solar. 
The ratio of total iron mass in both stars and gas 
to the total baryonic mass within 
$r_{sh}(t_n)$ corresponds to an iron 
abundance of $\sim 0.4$ in solar units, again very
similar to cluster values. 
This apparent agreement is likely to be just a coincidence, 
however.
The important distinction between our model for 
NGC 4472 based on a Salpeter $\eta_{II}$ and higher 
values of $\eta_{II}$ required to understand 
cluster ICM abundances is in the 
much lower total gas mass fraction currently present 
in galaxies as compared to clusters.
Within any radius of interest out to the current accretion 
shock radius, the ratio of gas mass to total baryonic mass 
in our standard model 
is much lower than that found in rich clusters.
If the ICM in rich clusters contains additional, 
unenriched primeval gas at larger radii, 
the enrichment of cluster gas by ellipticals similar 
to our model for NGC 4472 is 
even more dramatically inadequate. 
Therefore, our models support the contention made 
by Renzini (1997) and David (1997) that 
it is impossible to build presently observed rich clusters 
by merging 
hot gas in group-dominant massive ellipticals 
having properties similar to 
our standard solution for NGC 4472.

We have made additional 
calculations using standard model parameters 
but assuming that the SNII energy and enrichment occur 
within radii that are not equal to $r_{sh}(t_*)$.
When the enrichment region is smaller than $r_{sh}(t_*)$, the 
plateau region is smaller but its iron abundance is increased 
so that it is no longer possible to match the current negative 
iron abundance gradients observed in NGC 4472 
and other similar massive ellipticals. 
To correct for this excess iron, 
the value of $\eta_{II}$ would need to 
be less than $\eta_{std}$, diverging further from 
typical cluster values.  
If the size of the enrichment radius is much larger than 
$r_{sh}(t_*)$,
then the dynamical time for other stars spatially associated 
with the enriching SNII to arrive at the outer radius of 
the currently observed stellar system in NGC 4472 
exceeds the time available $t_n - t_*$. 
For example, stars that form at $t_*$ beyond 
$r \approx 700$ kpc in the standard solution can 
never enter the stellar part of the 
galaxy ($r \lta 100$ kpc) by the present time $t_n = 13$ Gyrs; 
any SNII enrichment beyond this radius is unrelated to 
stars within NGC 4472.
In any case, even if we use a SNII energy and 
enrichment radius equal to the maximum allowed size at $t_*$,
$\sim 700$ kpc, the 
current abundances of iron and silicon are still too large 
if the cluster value of $\eta_{II}$ is used.


\subsection{The Cluster Model}

To emphasize the disparity between optimum values 
for the specific number of SNII 
in rich clusters and in single bright 
ellipticals, we briefly discuss a ``cluster'' model 
for NGC 4472 in which 
$\eta_{II}$ has a larger value appropriate for rich clusters.
Following our previous discussion in \S 2, we assume 
$\eta_{II} = 2.36 \eta_{std}$ with $\beta = 0.55$ and 
a slightly larger mean SNII iron yield, 
$\langle y_{Fe,II} \rangle = 0.158$.
All remaining variables and the gas dynamical solution 
are identical to those in the standard model previously discussed. 
Larger $\eta_{II}$ corresponds to a lower heating efficiency 
$\epsilon_{sn}$, keeping $E_{II}$ 
and the associated gas dynamics unchanged from the standard 
solution.

After evolving to the current time, the iron and silicon 
distributions in this model, shown in Figure 6, 
are clearly unable to account for gas abundances observed in 
NGC 4472.
Iron and silicon within the galaxy 
are dominated by the inflow of 
SNII-enriched gas from outside the stellar system.
The total iron abundance within the optical galaxy 
($r \lta 100$ kpc) 
is much higher than observed abundances and 
its radial gradient is positive throughout the galaxy.
This model is an example of the sensitivity of $z_{Fe}(r)$ 
to the parameter $\eta_{II}$.
Increasing $\eta_{II}$ by a factor of 2.36 causes the gas 
abundance in the plateau region to increase to 3.35, 
almost 22 times greater than the abundance in this region 
in the standard solution.
As before most of the iron mass in the gas is contained in 
this distant plateau region (see Figure 5).

We explored several variants of the ``cluster'' model
in an attempt to reconcile cluster values of $\eta_{II}$ and 
the enrichment history of large ellipticals.
The high abundances in the ``plateau'' region just beyond 
100 kpc can be reduced if the radius of SNII enrichment at 
time $t_*$ is increased or if this gas is (rather arbitrarily) 
mixed and diluted with primordial gas at larger radii.
However, as discussed above, such models are 
unrealistic since 
the dynamical times at $\sim 700$ kpc for either 
of these variant models is too long. 
Low mass stars accompanying metal-producing 
SNII at these radii 
will not have joined the galaxy by the present time.
We have also sought solutions in which the galaxy formation 
time $t_*$ is 3 or 4 Gyrs in an attempt to lower the abundances 
by mixing SNII ejecta with larger masses of unprocessed gas 
inside the accretion shock at these later times, 
but unrealistically high iron abundances are still present. 

We are faced with the unexpected result that the 
amount of metals required to fit abundances in rich clusters 
are incompatible with those observed in individual ellipticals.
This is very curious since bulge dominated E and S0 galaxies 
are generally thought to be the primary source of metals 
in the ICM of rich clusters.

\subsection{The Ia Model}

We now explore the possibility that SNIa are important
contributors 
to the iron abundance in rich clusters
and that bulge dominated galaxies are the primary 
sites of these SNIa explosions.
This can be accomplished with the 
``Ia'' model for NGC4472 for which $n_{Ia} = 0.011$,
corresponding to $F_{Ia} = 0.22$ and 
$F_{Ia,g} = 0.25$ in Figure 1.
To achieve this high SNIa rate we choose 
$SNu(t_{n}) = 0.1$ and $p = 1.5$;
this is the same slope as that of Ciotti et al. (1991), but our 
$SNu(t_{n})$ is only 70\% of their value. 
All other parameters are identical to those in the standard model.

Since the energy deposited by SNIa in this model 
exceeds that of the standard model, 
the hydrodynamic solution is also different.
The gas density and temperature for this model after
time $t_n = 13$ Gyrs are shown with dashed lines in Figure 3.
The agreement with the observed gas density in NGC 4472 is good,
but the gas temperature has a
minimum near 60 kpc that is not observed.
This curious feature is a long-lived relic resulting from 
gas flows driven by SNIa and SNII heating at times just after 
time $t_*$.
In earlier theoretical models for the evolution of 
hot gas in ellipticals in which stars are 
assumed to be the only 
source of interstellar gas, 
such a large energy deposition by SNIa would 
have driven a strong galactic wind.
For the model we discuss here, where the young 
galaxy is surrounded by cosmic gas converging toward the 
overdensity perturbation 
(secondary infall), early SNIa-driven winds are suppressed 
by the inertia of this ambient gas and only 
a modest outward redistribution of the gas occurs.

Apart from the small temperature minimum in Figure 3, 
the general trend
in the gas temperature for this model is a reasonably good
fit to gas temperatures observed in NGC 4472.
However, the current iron and silicon 
abundances in the hot gas for this model, 
illustrated in Figure 7, 
are clearly at variance with X-ray observations. 
In this SNIa-dominant 
model the interstellar enrichment of iron and silicon is 
caused almost exclusively by SNIa, 
raising the iron abundance to $\gta 6$ times 
solar throughout the optical galaxy.
The shallow positive gradient $d z_{Fe}/dr > 0$ 
for $r \lta 100$ kpc is due to a small residual gaseous outflow 
caused by the large energy released by SNIa inside the galaxy.
The global value of the iron abundance and its radial 
gradient are both too large.
This model is rather similar to 
the earlier result of Loewenstein \& Mathews 
(1991) in their study of 
simpler galactic cooling flows without cosmological secondary infall; 
they also found iron abundances far in excess of those observed 
when SNIa are important contributors to the overall galactic 
enrichment. 
It is also significant that the silicon abundance in this model 
is much higher than the global value observed in the 
hot interstellar gas of NGC 4472,
$\langle z_{Si} \rangle = 0.5$ solar.
Most of the excess iron and silicon 
inside the galaxy in this model comes from Type Ia supernovae 
with a smaller contribution from stellar mass loss.
Because of the low-level, subsonic outflow driven by SNIa, 
iron and silicon introduced into the gas 
by SNII at $t_*$ are unable to enter the interstellar gas 
interior to $\sim 100$ kpc.

We have also explored 
additional high $n_{Ia}$ models in which the SNIa varies 
like a step function rather than a power law.
By concentrating most of the SNIa energy at early times,
 we hoped 
that the large amount of metals produced then would be expelled 
in a powerful galactic wind.
If so, the low gas abundances observed today within $\sim 100$ kpc 
might be made consistent with large SNIa enrichment at 
early times.
For example, we considered a model with $n_{Ia}=0.05$, with most
of the SNIa exploding at very early times: SNu = 24.86 for $2<t<4$ Gyr,
and SNu = 0.03 for $t > 4$ Gyr. However, with this sort of model 
the gas dynamics were altered so that 
we were unable to match $n(r)$ and $T(r)$ 
currently observed in NGC 4472. 
The iron and silicon abundance distribution are also strange in 
these models at the present time, with an
enormous metal enrichment occurring 
just outside of the stellar part of the galaxy.

\section{Conclusions and Final Remarks}

When we began this study of the enrichment history 
of massive ellipticals and rich clusters,
we quite mistakenly anticipated that it would be possible to 
find a common set of parameters that would 
account for evolutionary enrichment on 
both galactic and cluster scales.
Gas phase iron abundances in bright ellipticals range from 
$0.1 - 1$ solar, with NGC 4472 lying near the upper limit 
of this range, but the average may not differ 
greatly from the mean global iron abundance in 
rich clusters, $\sim 0.4$.
Rich clusters typically have higher gas fractions 
than massive ellipticals, but this might result in part from 
supernova-driven outflows from the shallower 
potentials of member ellipticals. 
Unlike the observational information available 
for hot gas in group-dominant
ellipticals, our computational models 
are closed boxes since we can computationally follow 
gas-phase supernova 
enrichment products into regions of very low 
gas density and emissivity where observations are difficult
or impossible. 
Perhaps, we supposed, the large number of past supernovae required 
to produce metal abundances in rich clusters were 
at one time also present in giant ellipticals but some of 
these metals was dispersed by galactic 
winds into the low density environment, 
escaping observational detection.
If all of this were true, it might be possible to merge small 
elliptical-dominated groups together with their 
wind-enriched gaseous environments and construct rich clusters.
In developing this hypothesis, 
we adopted the commonly held assumption that virtually all metal 
enrichment in rich clusters is due to the stellar 
spheroidal component, i.e. in E and S0 member galaxies. 

As discussed in our recent paper (Brighenti \& Mathews 1998b),
the observed hot gas density and temperature distributions
in any massive elliptical can be fit with a range of gas dynamical 
models having different numbers of Type II supernovae $N_{II}$, 
represented here by the SNII number per unit stellar mass, 
$\eta_{II} = N_{II}/M_*$, or the total energy released 
by all SNII, $E_{II} = \epsilon_{sn} \eta_{II} M_* E_{sn}$.
For a given set of cosmological parameters, hydrodynamical models 
for the evolution of hot 
interstellar gas are degenerate in the sense that 
a range of values of $\eta_{II}$ (associated with different IMFs) 
and therefore $E_{II}$ can produce similar interstellar 
density and temperature distributions at the present time 
provided the time of galaxy formation $t_*$ is also adjusted. 
If the galaxy is assumed to form at a later time, more gas is 
available near the central overdense region and a larger value of 
$E_{II}$ generates approximately the same 
specific thermal energy; 
the subsequent evolution of the gas arrives at similar 
$n(r)$ and $T(r)$ distributions at the present time.
This approximate degeneracy depends only weakly on our assumption 
that the SNII energy is deposited within the accretion shock 
radius at time $t_*$, $r_{sh}(t_*)$.
As we have discussed already, the extent of the 
SNII enrichment region $r_{II}$ in massive ellipticals 
cannot be greatly different from $r_{sh}(t_*)$.
If $r_{II} < r_{sh}(t_*)$, 
then even lower values of $\eta_{II}$ are required 
to reproduce abundances observed today, increasing 
the disagreement with the higher cluster 
values for $\eta_{II}$. 
If $r_{II} > r_{sh}(t_*)$, then
the dynamical time also increases and  
low mass stars associated with the SNII enrichment would not have 
enough time to enter the massive optical galaxy 
observed today.

The important parameters that describe our evolutionary models 
are (i) the total energy released by Type II supernovae,
$E_{II} = \epsilon_{sn} \eta_{II} M_* E_{sn}$ 
which must be adjusted to achieve agreement with 
$n(r)$ and $T(r)$ in the hot interstellar gas,
(ii) $t_*$ the time of star formation, 
(iii) $\eta_{II}$ which controls most of the gas enrichment, 
and (iv) the SNIa rate SNu$(t)$ that supplies much of the 
interstellar iron in the central galaxy, $r \lta 50$ kpc.
In seeking satisfactory gas dynamical models with no 
attention to abundances, the main parameters are 
$E_{II}$ and $t_*$.
We have found that models for the large elliptical
NGC 4472 agree with $n(r)$ and $T(r)$ {\it and} the 
abundance distribution $z_{Fe}(r)$ 
if stars form at $t_* = 2$ Gyrs with
a Salpeter IMF for which $\eta_{II} = \eta_{std} \equiv 6.81
\times 10^{-3}$ assuming a heating efficiency $\epsilon_{sn} = 1$.
If we wish to increase $\eta_{II}$
and therefore $E_{II}$ toward the global values
indicated by rich cluster abundances, $\sim 2 - 3 \eta_{std}$,
$t_*$ must be increased significantly.
However, such an increased  $t_*$ is not possible 
since observations indicate that many luminous 
elliptical galaxies are very old: 
the color-magnitude diagram (Bower, Lucey, \& Ellis 1992),
the small scatter in the Mg$_2 - \sigma$ relation 
(Bender, Burstein, \& Faber 1993), 
passive evolution of the fundamental plane 
(van Dokkum \& Franx 1996), etc.
With $t_*$ constrained to values $\lta 2$ Gyrs, 
and $E_{II}$ constrained for a match currently observed 
$n(r)$ and $T(r)$ distributions,
$\eta_{II}$ can be increased toward 
the cluster value ($\sim 2.6 \eta_{std}$) only by decreasing 
the gas heating efficiency $\epsilon_{sn}$ with 
$E_{II}$ and $t_*$ held fixed.
This is the procedure we have 
considered with the ``cluster'' or high-$\eta_{II}$ model 
for NGC 4472. 
However, as $\eta_{II}$ is increased above the 
standard Salpeter value, the metal enrichment 
of the interstellar gas at the present time quickly 
rises far above abundances observed in NGC 4472.
The global iron and silicon abundances and the
negative radial iron gradient $d z_{Fe}/dr$ 
in NGC 4472 (and other large
ellipticals) can be matched only with a very narrow range of 
$\eta_{II}$, all lower than cluster values.
Similar restrictions also apply to the total number of  
Type Ia supernovae which we assume are also more frequent in 
the early universe. 

We arrive at the following conclusions:

(1) Type II supernova production with a Salpeter IMF is sufficient
to explain all important hot gas observations in bright
ellipticals: the radial distribution of density, temperature and
iron abundance as well as the global iron and silicon abundances.
A small additional contribution of iron and silicon from 
Type Ia supernovae, consistent with currently observed SNIa 
rates in elliptical galaxies, can also be present.
Unlike the enrichment history of our Galaxy,
where iron production by Type II and Type Ia supernovae
have been comparable, in ellipticals the total iron and silicon 
enrichment by Type Ia supernovae must be much 
less than that from SNII.

(2) The gas dynamic and enrichment history of massive 
ellipticals establishes upper limits on the total 
number of both Type II and Type Ia supernovae events. 
Large numbers of Type II supernovae associated 
for example with flat IMFs, often invoked to account for the 
iron observed in rich clusters, produce too much gas-phase 
iron and silicon in models of single massive ellipticals. 
If the total number of Type Ia supernovae is large enough 
to dynamically influence the interstellar gas 
in these galaxy models, the 
current metal enrichment greatly exceeds 
the gas phase iron and silicon abundances observed.

(3) For fixed total SNII energy $E_{II}$, the iron and silicon 
enrichment is very sensitive to the specific number of Type II 
supernovae $\eta_{II}$ involved. 
This sensitivity may indicate that our galaxy enrichment 
model is oversimplified 
or it may be related to the wide spread in 
gas phase abundances observed 
in ellipticals and group-dominant ellipticals 
(see Fig. 6 of Renzini 1997).

(4) 
We agree with David (1997) and Renzini (1997) that
it is not possible to form presently observed rich clusters
by simply combining stars and hot gas from
a large number of presently observed elliptical-dominated groups.

(5) In view of our model calculations for the evolution 
and enrichment of interstellar gas in massive ellipticals,
{\it neither} of the currently-discussed explanations of 
cluster metal abundances is particularly attractive:
(i) IMFs flatter than Salpeter, and 
(ii) substantial iron enrichment from Type Ia supernovae,
as in our Galaxy.

These conclusions could be changed if some of the 
theoretical and observational parameters 
we have used are in error by large amounts -- for example if 
observed metal abundances in the hot interstellar gas 
of ellipticals are underestimated by factors of 
5 or more -- but this seems unlikely. 

Alternatively, 
the different enrichment histories of large 
ellipticals and rich clusters could be reconciled 
if the cluster gas is enriched by another source not involving 
massive elliptical galaxies.
Arnaud et al. (1992) show that the total 
mass of hot gas in clusters correlates nicely with the total 
optical luminosity of E and S0 galaxies 
within 3 Mpc of the cluster centers, 
but there is no similar correlation 
with the total luminosity of the spiral galaxy component.
Since global gas phase iron abundances are fairly constant 
among clusters, a correlation with gas mass is also 
a correlation with the total mass of iron, 
most of which is in the gas phase.
Arnaud et al. (1992) therefore concluded that 
E and S0 galaxies are the only significant 
sources of metal enrichment in the hot cluster gas.

However, cluster enrichment may be more complicated than this.
For example, the role of disk galaxies in ICM enrichment 
may be more important than is currently thought.
Most low redshift rich clusters 
contain about twice as many S0 galaxies as ellipticals 
(Dressler 1980). 
The fractions of S0 and spiral galaxies in clusters 
are inversely related while the ratio of the number of 
S0 plus spiral galaxies to the number of ellipticals 
is rather constant.
This suggests 
that S0 galaxies may be descended from spirals.
This is supported by the observation that
richer clusters tend to have higher S0/spiral ratios.
Even more relevant, cluster observations at $z \sim 0.5$ 
reported by Dressler et al. (1997)
show that the fractional content of ellipticals is similar
to nearby clusters but the S0/spiral ratio is very much lower.
This can be naturally explained if S0 galaxies in rich clusters 
were spirals in their previous lives before 
interacting with other cluster galaxies and the ICM.
When stripped of their interstellar gas by the ICM, 
S0 galaxies could have enriched the ICM, 
perhaps with a strong SNIa abundance signature.
If the ICM enrichment were due solely to this process, however, 
the iron abundance in the ICM should 
decrease systematically with redshift. 
Mushotzky \& Loewenstein (1997) 
find no significant change in the 
global iron abundance in clusters out to redshifts $z \sim 0.3$,
although there is considerable real scatter in the data that 
may have masked such a trend.
While the transition of spirals into S0 galaxies is 
appealing as a possible explanation for the large specific 
supernova rates in clusters, morphological classification 
at large redshift is a difficult art and subject to error.
In similar studies of distant clusters 
Stanford et al. (1997) and Andreon (1998) find no depletion 
of S0 galaxies at high redshift.

Another less conventional cluster enrichment 
possibility is that some of the cluster gas metals 
has come from outflows from dwarf galaxies. 
At least one author (Trentham 1994) has claimed that 
dwarf galaxies can supply all of the gas and metal enrichment 
observed in rich clusters. 
While few would agree with this extreme hypothesis, 
some fractional contribution from dwarf galaxies to the ICM 
enrichment cannot be entirely dismissed.
>From studies of intergalactic absorption lines in quasar spectra,
Cowie \& Songaila (1998) 
and Lu et al. (1998) find evidence of an average carbon enrichment 
of $\sim 0.003$ solar in the intergalactic medium at 
high redshifts $z \sim 3$, even in low density voids. 
Although this value is 100 times lower than abundances
in rich clusters, it does suggest the presence of 
an alternative source of metal enrichment.
According to Cowie \& Songaila, 
``early generations of small galaxies might 
be much more efficient at ejecting heavy
elements ... than has previously been thought.''
Lu et al. maintain that the variation of the carbon abundance 
with column (and physical) 
density in L$\alpha$ clouds rules out the possibility 
of (uniform) metal contamination of the 
intergalactic medium by population III objects.

In another intriguing recent observation with ASCA, 
Hattori, et al. (1997) 
have discovered an X-ray cluster at 
a very high red shift $z = 0.92$ 
having luminosity $L_x = 8 \times 10^{44}$ erg s$^{-1}$. 
It is most remarkable that this cluster is optically dark.
After several 
sensitive optical searches only the central CD galaxy 
has been found at this redshift.
Other cluster member galaxies of 
average luminosity, if present, should have been observed 
and were not.
This cluster is similar in all respects to those at lower 
redshifts ($z \lta 0.4$) except that its mass to light ratio,  
$\sim 3000$, is about ten times that of less distant clusters. 
Given the apparent absence of 
prominent cluster galaxies, it is most astonishing that 
the iron abundance in the hot cluster gas ($kT = 8.6$ keV) is 
$z_{Fe} = 1.7^{+1.25}_{-0.74}$ in solar units.
One possible interpretation would be that the 
cluster gas is enriched by a multitude of low mass 
galaxies which has escaped optical detection. 
If dwarf galaxies created some of the metals in rich clusters, 
it is likely that this happened at high redshifts 
to account for the relative constancy of the iron abundance 
found by Mushotzky \& Loewenstein (1997) at 
red shifts $z \lta 0.3$.

Some of these alternative cluster enrichment hypotheses 
are subject to observational test.
We have already mentioned that there is a small, but real, 
scatter in the global iron abundance in cluster gas
(Mushotzky 1998).
Since the S0/spiral ratio also varies among clusters, 
a weak positive 
correlation of this ratio with gas phase abundance excess 
would be expected if S0 stripping is an important 
contributor to cluster gas enrichment.
Abundance inhomogeneities in the hot cluster gas may 
provide another clue to the origin of cluster metals.
If only a few bright E galaxies are responsible for 
most of the intracluster gas enrichment,
significant abundance inhomogeneities and radial gradients 
would be expected due to the limited previous orbital 
experience of these galaxies.
If the cluster gas was enriched by a larger number of 
dwarf galaxies (most of which may no longer be visible),
then abundance inhomogeneities will be much reduced.
Such studies may become possible as 
the spatial resolution of X-ray detectors improves. 

The possibility of galactic 
sources of cluster enrichment other than E and S0 galaxies 
has been considerably strengthened by recent 
ASCA observations of cluster iron and silicon abundances.
In a study of 40 clusters 
Fukazawa et al. (1998) find that the ICM silicon abundance 
increases with cluster richness as $kT$ increases from 
$\sim 2$ keV to $\sim 9$ keV,
but no appreciable change in the iron abundance is 
indicated over this range.
The explanation for this behavior is unclear at present, 
but it very definitely requires at least two sources 
of ICM enrichment, otherwise the iron to silicon 
ratio would be the same in all clusters.
For this range of cluster gas temperatures, 
$kT \gta 2$ keV, the constancy of global gas and iron masses 
relative to cluster ${\cal L}_B$ suggests 
closed box environments (Renzini 1997).
However, the sense of the silicon enrichment is 
opposite to that expected if more spiral galaxies 
in richer clusters (high $kT$) convert to S0s.
Nevertheless, additional sources of 
cluster gas metals may provide the most satisfactory 
resolution of the inconsistent enrichment histories of 
massive ellipticals and rich clusters that we have 
demonstrated here.

\acknowledgments

Studies of the evolution of hot gas in elliptical galaxies 
at UC Santa Cruz is supported by
NASA grant NAG 5-3060 and an NSF grant 
for which we are very grateful. In addition
FB is supported
in part by Grant ARS-96-70 from the Agenzia Spaziale Italiana.

\clearpage

\centerline{\bf APPENDIX}
\appendix

In this Appendix we briefly discuss 
the evolution of hot gaseous halos in NGC 4472 
in the context of two different cosmologies: an open universe 
with $\Omega = 0.3$ and a flat, low density universe 
with $\Omega = 0.3$ and $\Omega_{\Lambda} = 0.7$.
When $\Omega \neq 1$ the evolution of the dark matter can 
no longer be described by the self-similar solutions 
of Bertschinger (1985).
Instead, we adopt a simple top hat overdensity perturbation 
at some early time 
designed to concentrate the dark mass of NGC 4472 
by the current time. 
The dark matter is followed as a separate 
zero pressure fluid which builds 
a stationary NFW core 
that grows with time, conserving total dark mass.
For simplicity we assume the NFW profile for NGC 4472
is identical to that in the flat universe described 
in the main text. 

\section{Halo and 
ISM Evolution in an $\Omega = 0.3$ Universe}

Since all matter in an open universe is unbound, the 
initial perturbation must be large and non-linear 
in our simple model of galaxy formation. 
For the $\Omega = 0.3$ universe  
we introduced a top hat perturbation of 
overdensity $(1 + \delta) \rho(t_i)$ with 
$\delta = 5$ extending to $R_i = 43.5$ kpc at time
$t_i = 10^8$ yrs.
Baryons with mass density $\Omega_b = 0.05$ 
also participate in this 
top hat perturbation and by time $t_* = 1.5$ Gyrs enough
gas is concentrated to form the stars in NGC 4472 in the 
usual way described in the main text.
We chose $\eta_{II} = 0.3 \eta_{std}$ for the SNII number 
(taking $\epsilon_{sn} = 1$) and the Type Ia 
supernova rate is specified 
with parameters SNu$(t_n) = 0.03$ and $p = 1$.
In Figure 8 we show with a solid line the gas density 
and temperature distributions after time $t_n = 13$ Gyrs.
Although the agreement is not perfect, 
it is adequate 
to illustrate that the global evolution of 
extended hot halos in ellipticals 
in an open universe is quite similar 
to the solution for the flat $\Omega = 1$ 
universe discussed in the main text.

At 13 Gyrs the transition from the stationary 
NFW dark halo core 
to the secondary infall occurs at $\sim 200$ kpc, which 
is just beyond the outermost observations of the X-ray
image of NGC 4472.
Since the potential has a strong slope 
change at this radius, this feature at the outer dark core 
might be visible in the outer X-ray images of 
some massive ellipticals if the universe were open.
For open universe parameters, $t_*$ cannot be much larger 
than $\sim 1.5$ Gyr otherwise too much cold gas  
gathers in the perturbation potential. 
In the open cosmology the total baryon fraction is 
higher at large radii $\sim 0.16$, 
similar to rich clusters; although the final 
gas density is necessarily similar to that in 
the flat universe solution, 
the dark matter fraction is lower. 

We also computed the radial iron and silicon abundances in 
the gas at time $t_n = 13$ Gyrs and find a general agreement 
with observed abundances, although the fit could be improved 
with a slightly lower $\eta_{II}$.
Successful solutions for NGC 4472 in the context of an 
open universe are incompatible with 
large $\eta_{II} \approx 2.6 \eta_{std}$ required 
for global cluster enrichment.

\section{Halo and ISM Evolution in an $\Omega = 0.3$,
$\Omega_{\Lambda} = 0.7$ Universe}

Galaxy evolution in this cosmology was initiated 
with a top hat perturbation of radius 
$R_i = 3$ kpc and relative overdensity 
$\delta = 0.59$ at time $t_i = 10^6$ yrs.
For the supernova rates we take $\eta_{II} = \eta_{std}$ 
and SNu$(t_n) = 0.03$ with $p = 1$.
The stars in NGC 4472 are assumed to form at 
$t_* = 3$ Gyrs.
After 13 Gyrs the gas density and temperature 
profiles are shown 
with dashed lines in Figure 8.
The gas density profile is slightly too peaked but 
is otherwise similar to the observed 
variation.
The gas temperature exhibits a curious 
undulation near $r = 90$ kpc that is not apparent in the 
observations.
Nevertheless, the computed temperature profile 
for this model agrees with observed temperatures 
far better than profiles computed without
secondary infall (see Brighenti \& Mathews 1998b).

Our purpose in illustrating these results with alternate 
cosmologies is not to achieve the best 
possible fits to the current 
density and temperature distributions 
by careful adjustment of the parameters.
We only wish to demonstrate that such models are 
feasible and that the principal conclusions 
reached in this paper do not depend on the simple 
flat cosmology that we adopt.






\clearpage


\vskip.1in
\figcaption[aasabundfig1.ps]{
Solutions of Equations (4) - (6) for cluster enrichment 
parameters in terms of the number of Type Ia supernovae 
per unit $L_B$, $n_{Ia}$ (bottom x-axis) and 
the current Type Ia rate SNu$(t_n)$ in SNu units (top x-axis).
{\it Upper left:} Number of Type II supernova per stellar mass 
$\eta_{II}$;
{\it Upper right:} IMF-averaged iron yield for Type II 
supernovae $\langle y_{Fe,II} \rangle$;
{\it Lower left:} Fraction of metals from Type II supernovae 
in stars $f_*$.
{\it Solid lines:} correspond to no stellar enrichment by SNIa
($g_* = 0$); {\it Dashed lines} correspond to enrichment by
half of the SNIa ejecta ($g_* = 0.5$).
{\it Lower right:} Total fraction of iron produced by SNIa
$F_{Ia}$ ({\it dot-dashed lines}) and the total fraction of iron
in the gas phase produced by SNIa $F_{Ia,g}$
({\it dotted line}) when $g_* = 0.5$.
$F_{Ia}$ and $F_{Ia,g}$ for the $g_* = 0$ case
({\it solid line}).
\label{fig1}}

\vskip.1in
\figcaption[aasabundfig2.ps]{
{\it Upper panel:} Variation of the umber of Type II supernova per
stellar mass $\eta_{II}(x)$ with IMF slope $x$ for 
three pairs of upper and lower mass cutoffs:
{\it solid line:} ($m_u$, $m_{\ell}$) = (100, 0.08);
{\it long dashed line:} ($m_u$, $m_{\ell}$) = (40, 0.08);
{\it short dashed line:} ($m_u$, $m_{\ell}$) = (40, 0.3).
{\it Lower panel:} Variation of $\beta$,
the fraction of initial stellar 
mass expelled 11 Gyrs after a single burst of star formation.
The IMF notation is identical to that in the upper panel.
\label{fig2}}

\vskip.1in
\figcaption[aasabundfig3.ps]{
{\it Upper panel:} Current 
gas density profile of the standard model
({\it solid line}) and the cluster model ({\it dashed line})
compared with observations of NGC 4472.
Filled circles are
NGC 4472 gas densities observed with
{\it Einstein} HRI (Trinchieri, Fabbiano \& Canizares 1986) 
and open circles are densities determined with 
ROSAT HRI and PSPC (Irwin \& Sarazin 1996).
{\it Lower panel:} Current 
gas temperature profile of the standard model
({\it solid line}) and the cluster model ({\it dashed line})
compared with ROSAT 
gas temperature observations of NGC 4472 
(Irwin \& Sarazin 1996) shown with errorbars.
\label{fig3}}

\vskip.1in
\figcaption[aasabundfig4.ps]{
Radial variation of iron ({\it upper panel}) and 
silicon ({\it lower panel}) abundances for the standard model.
The ASCA iron observations (from Matsushita 1997) 
are with the SIS detector ({\it filled circles}) 
and the GIS detector ({\it open circles}).
{\it Solid lines:} total abundances;
{\it Long dashed lines:} abundance contribution from SNII;
{\it Short dashed lines:} abundance contribution from SNIa;
{\it Dotted lines:} abundance contribution from stellar outflow.
\label{fig4}}

\vskip.1in
\figcaption[aasabundfig5.ps]{
Cumulative mass of iron in the hot interstellar gas 
at time $t_n = 13$ Gyrs as a function of galactic radius
for the standard model ({\it solid line}),
the cluster model ({\it dashed line}),
and the Ia model ({\it dotted line}).
\label{fig5}}

\vskip.1in
\figcaption[aasabundfig6.ps]{
Radial variation of iron ({\it upper panel}) and
silicon ({\it lower panel}) abundances for the 
``cluster'' model.
Notation is identical to that in Figure 4.
\label{fig6}}

\vskip.1in
\figcaption[aasabundfig7.ps]{
Radial variation of iron ({\it upper panel}) and
silicon ({\it lower panel}) abundances for the
``Ia'' model.
Notation is identical to that in Figure 4.
\label{fig7}}

\appendix

\vskip.1in
\figcaption[aasabundfigA1.ps]{
Current gas density ({\it upper panel}) 
and temperature ({\it lower panel}) profiles 
for the open universe ($\Omega = 0.3$) (solid lines) and 
the low density flat universe
($\Omega = 0.3$, $\Omega_{\Lambda} = 0.7$) (dashed lines).
plotted against observations of NGC 4472 
described in Figure 3.
\label{figA1}}

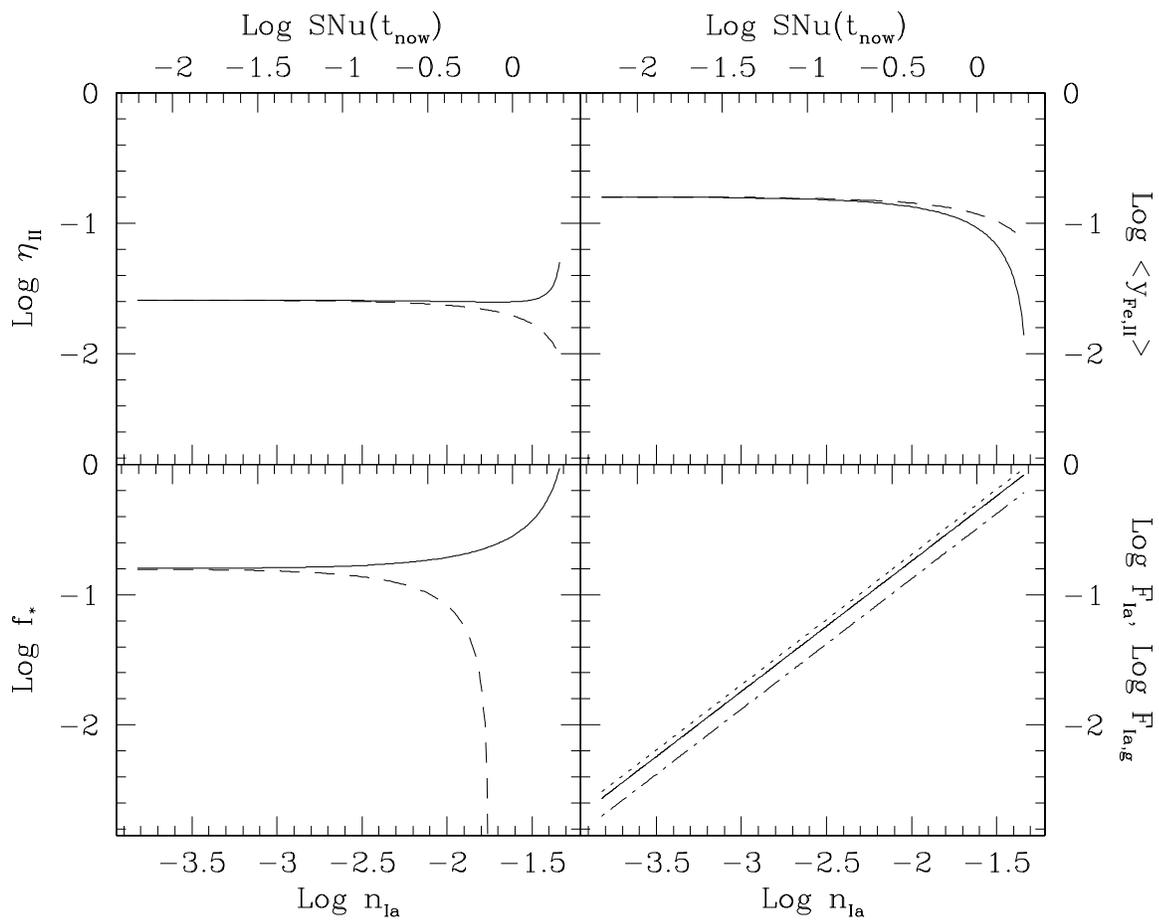
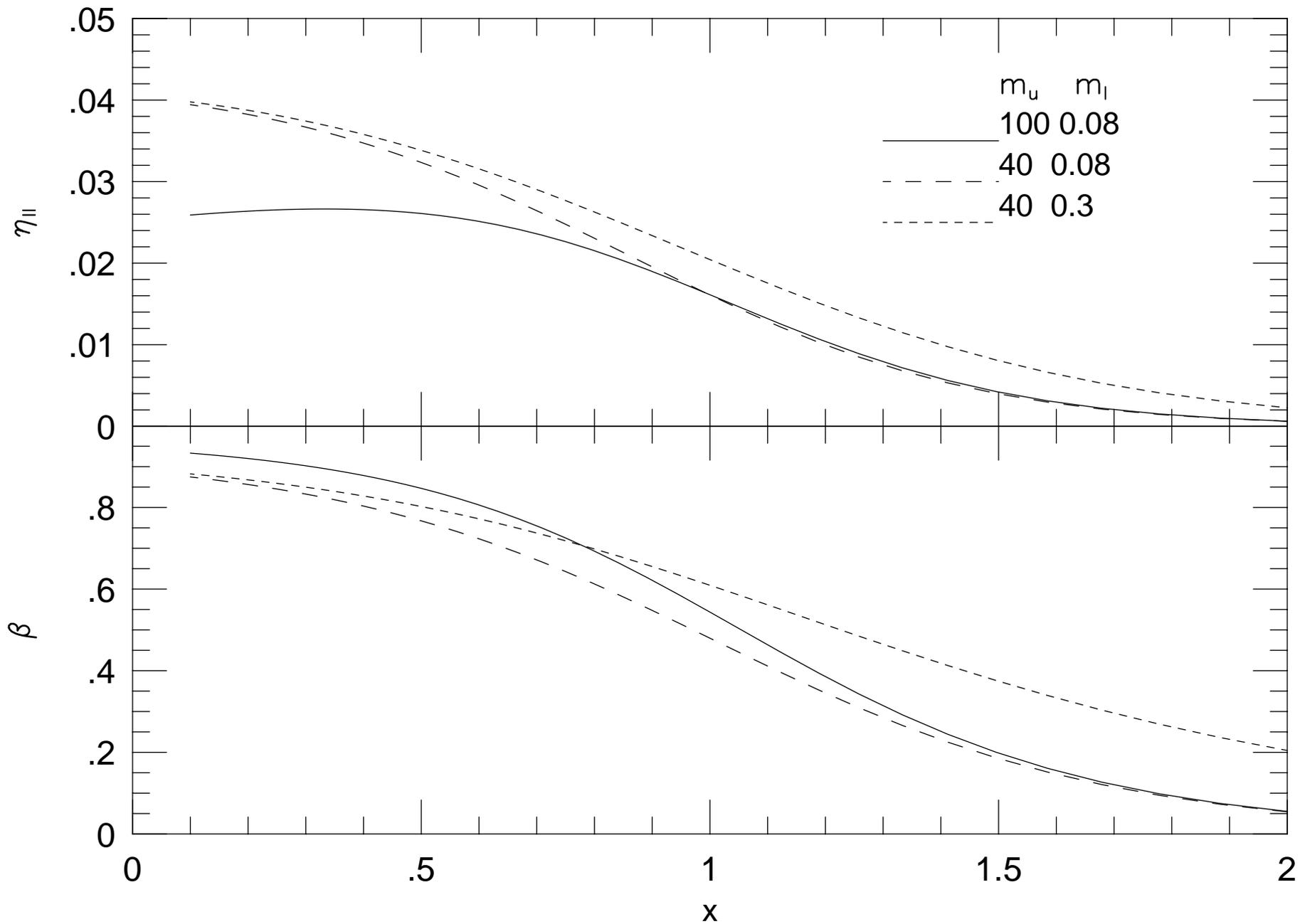
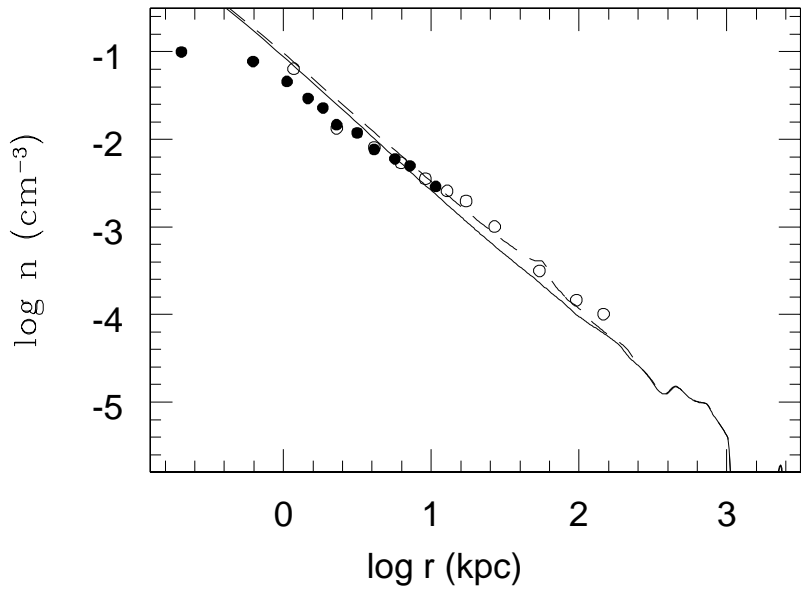
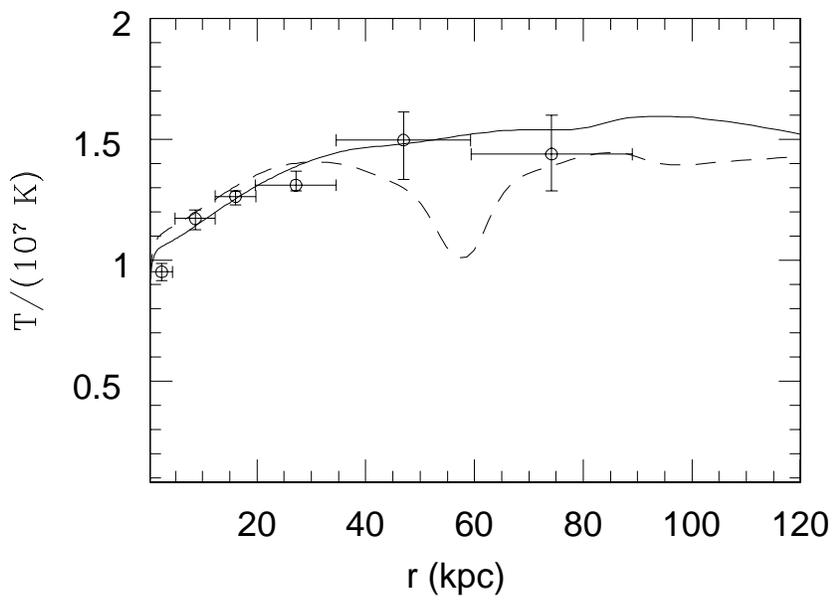
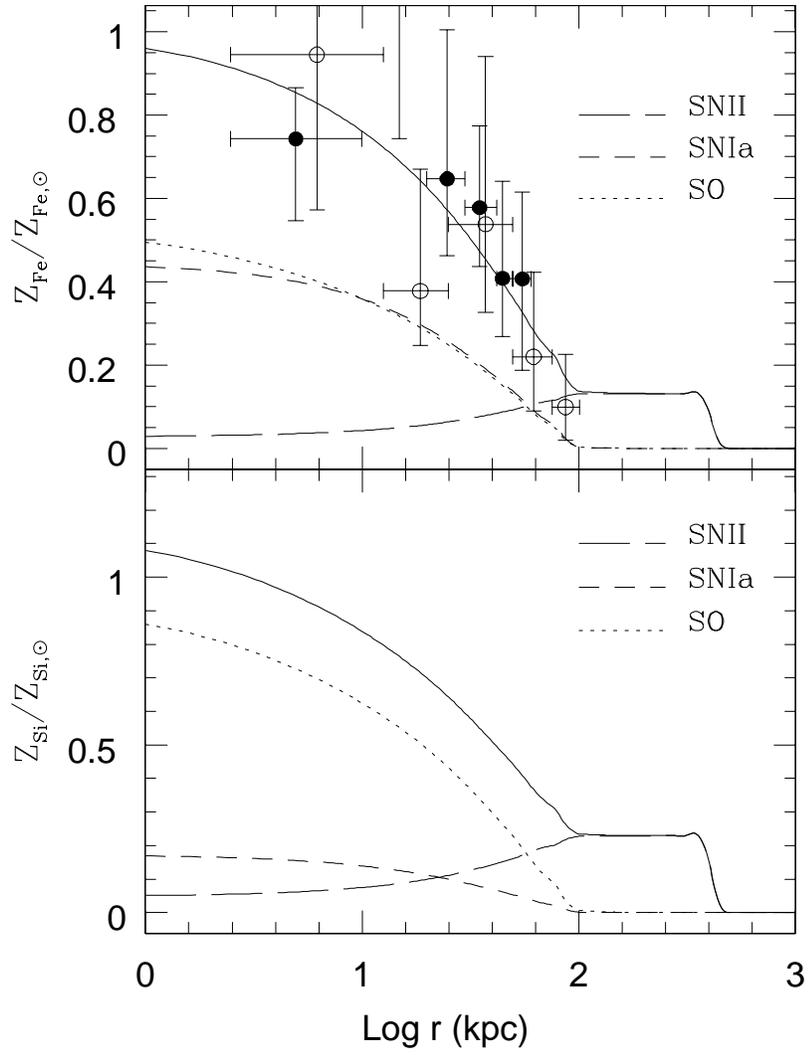
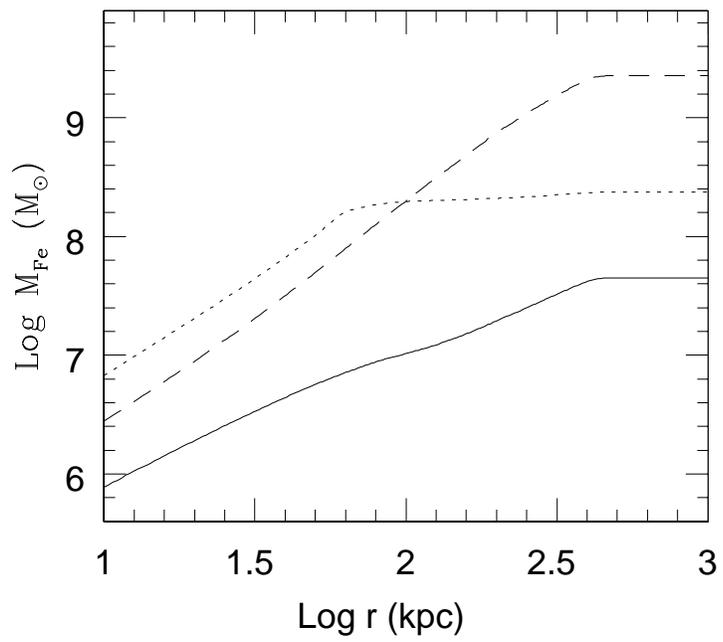
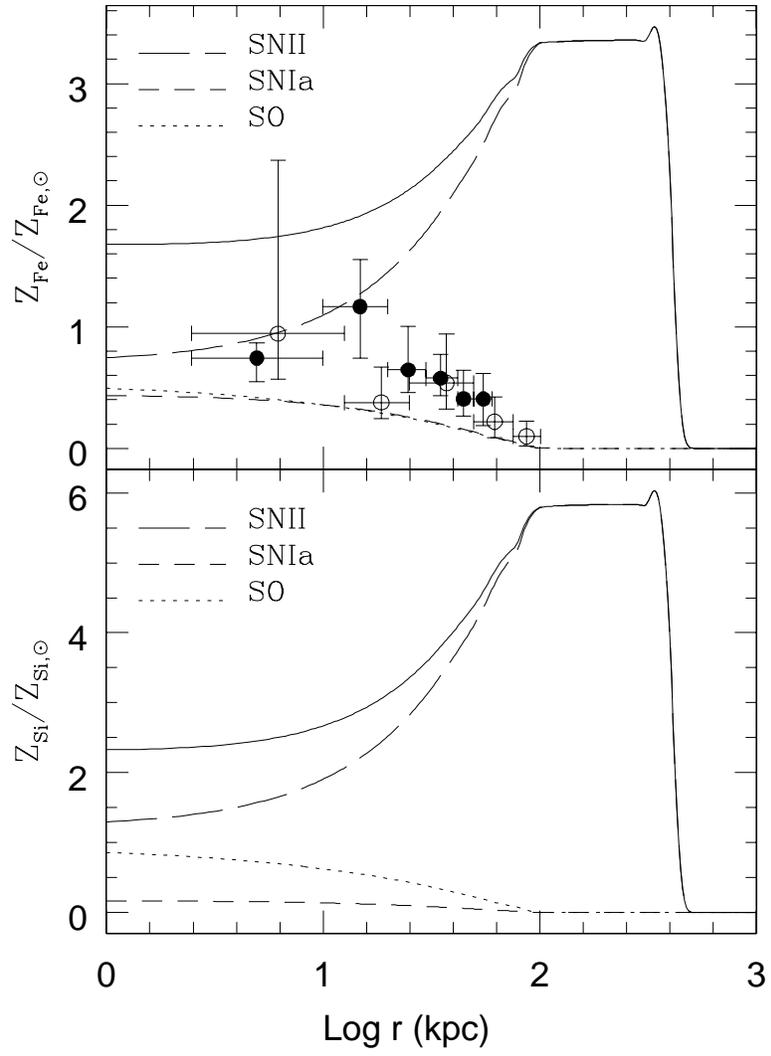
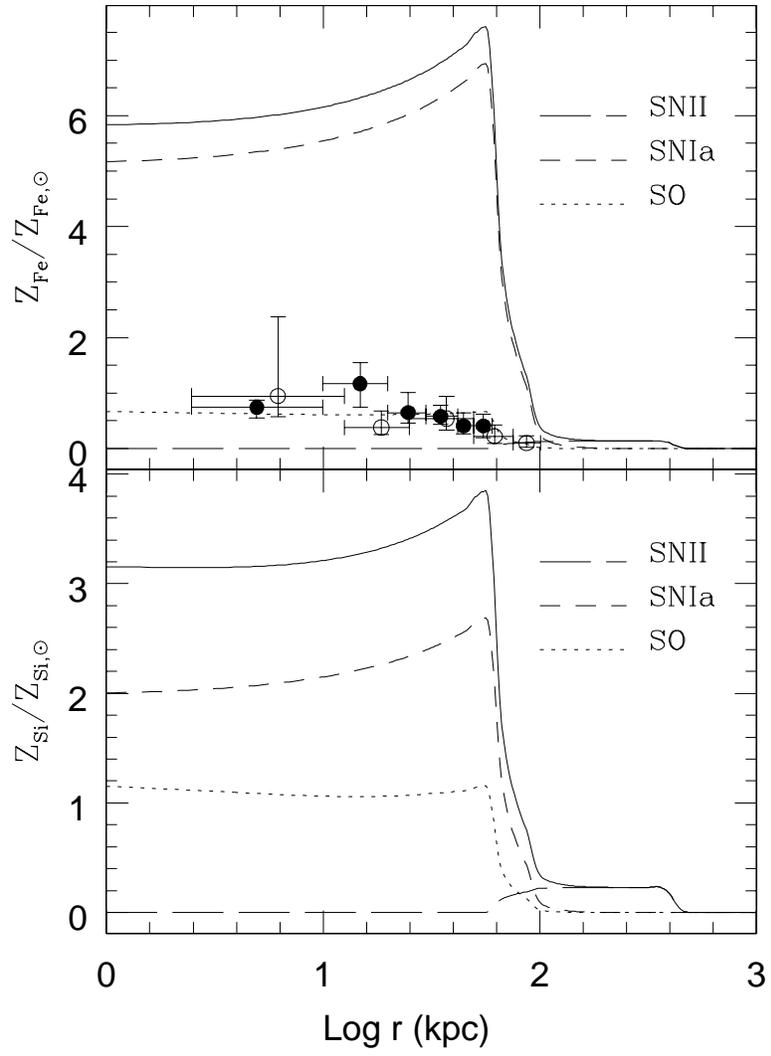
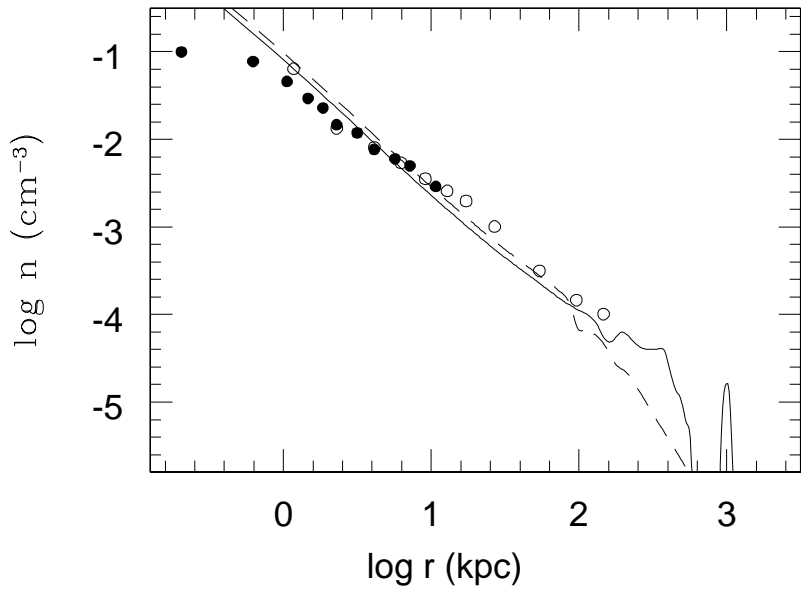
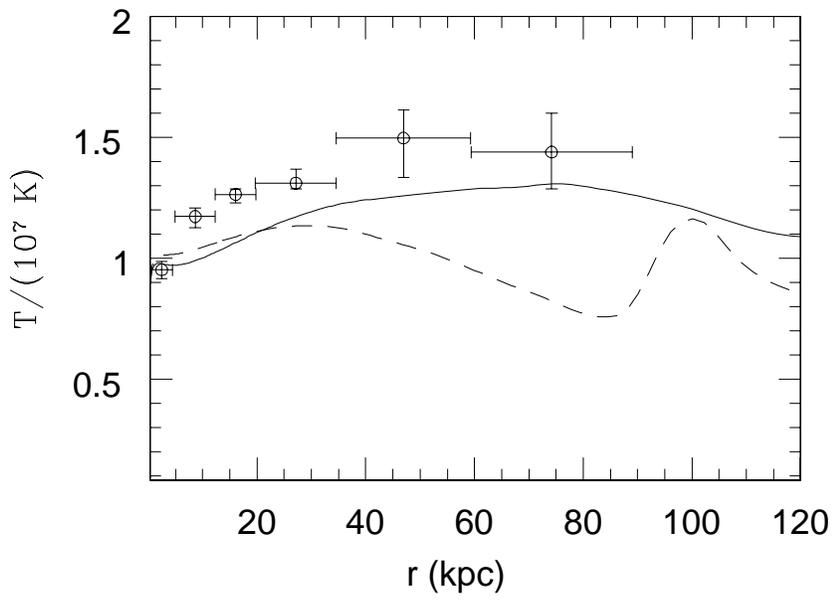
\end{document}